%% file: main.tex
\documentclass{elsarticle}

\usepackage[left=2cm, right=2cm, top=2.5cm]{geometry}

\usepackage{natbib}

\usepackage{graphicx,import}
\graphicspath{{./images/}{./../images/}}
\usepackage{pgf}
\usepackage{subfiles}
\usepackage{subcaption}
\usepackage[lcgreekalpha]{stix}
\usepackage{bm}
\usepackage{amsmath,mathtools}
\usepackage{tabularx}
\usepackage{multirow}

\usepackage[utf8]{inputenc}
\usepackage[T1]{fontenc}
\usepackage{textcomp}

\DeclareUnicodeCharacter{2212}{\textminus}

\usepackage[parfill]{parskip}

\usepackage{color, soul}

\newcommand\inputpgf[2]{{
\let\pgfimageWithoutPath\pgfimage
\renewcommand{\pgfimage}[2][]{\pgfimageWithoutPath[##1]{#1/##2}}
\input{#1/#2}
}}

\usepackage{setspace}
\usepackage{acro}
\DeclareAcronym{wc}{short=WC, long=weakly compressible}
\DeclareAcronym{ic}{short=IC, long=incompressible}
\DeclareAcronym{ft}{short=FT, long=front tracking}
\DeclareAcronym{fd}{short=FD, long=finite difference}
\DeclareAcronym{mlm}{short=MLM, long=meshless Lagrangian method}
\DeclareAcronym{gfd}{short=GFD, long=generalized finite difference}
\DeclareAcronym{sph}{short=SPH, long=smoothed particle hydrodynamics}
\DeclareAcronym{mps}{short=MPS, long=moving particle semi-implicit}
\DeclareAcronym{nse}{short=NSE, long=Navier-Stokes equations}
\DeclareAcronym{pde}{short=PDE, long=partial differential equation}
\DeclareAcronym{dem}{short=DEM, long=discrete element method}
\DeclareAcronym{ppe}{short=PPE, long=pressure Poisson equation}
\DeclareAcronym{gpu}{short=GPU, long=graphics processing unit}
\DeclareAcronym{les}{short=LES, long=large eddy simulation}
\DeclareAcronym{evm}{short=EVM, long=eddy viscosity model}
\DeclareAcronym{csf}{short=CSF, long=continuous surface force}
\DeclareAcronym{fvm}{short=FVM, long=finite volume method}
\DeclareAcronym{rmse}{short=RMSE, long=root mean squared error}
\DeclareAcronym{bicg}{short=BiCGSTAB, long=bi-conjugate gradient stabilized}

\newcommand{\gfd}[1]{\left<#1\right>}
\newcommand{\md}[1]{\mathbf{#1}}
\newcommand{\g}[1]{\nabla #1}

\def \divT {\gfd{\g\cdot\mathbf{\tau}}}
\def \mixL {\gfd{\mu\g^2\mathbf{u}}}

\begin{document}
\begin{frontmatter}
    \title{On the momentum diffusion over multiphase surfaces with meshless methods}
    \cortext[cor]{Corresponding author}
    \address[tuks]{Department of Mechanical and Aeronautical Engineering,
    University of Pretoria, South Africa}
    \address[imt]{IMT Lille Douai, Univ. Lille, EA 4515 - LGCgE Laboratoire de
    Génie Civil et géoEnvironnement, CERI Matériaux et Procédés, F-59000 Lille,
    France}

    \author[tuks]{Johannes C. Joubert\corref{cor}}
    \ead{jcjoubert.online@gmail.com}
    \author[tuks]{Daniel N. Wilke}
    \author[imt]{Patrick Pizette}

\begin{abstract}
This work investigates the effects of the choice of momentum diffusion operator
on the evolution of multiphase fluid systems resolved with Meshless Lagrangian
Methods (MLM). Specifically, the effects of a non-zero viscosity gradient at
multiphase interfaces are explored. This work shows that both the typical
Smoothed Particle Hydrodynamics (SPH) and Generalized Finite Difference (GFD)
diffusion operators under-predict the shear divergence at multiphase
interfaces.  Furthermore, it was shown that larger viscosity ratios increase
the significance of this behavior. A multiphase GFD scheme is proposed that
makes use of a computationally efficient diffusion operator that accounts for
the effects arising from the jump discontinuity in viscosity. This scheme is
used to simulate a 3D bubble submerged in a heavier fluid with a density ratio
of $2:1$ and a dynamic viscosity ratio of $100:1$. When comparing the effects
of momentum diffusion operators, a discrepancy of 57.2\% was observed in the
bubble's ascent velocity.
\end{abstract}

\end{frontmatter}

\section{Introduction}
The prediction of multiphase flow phenomena is a widespread problem in various
physical and engineering environments with applications in fields such as oil
and gas production, power generation, and chemical processing. Recently,
\Acp{mlm} have become a popular alternative to mesh-based approaches for
simulating multiphase flow due to them avoiding many of the challenges
introduced when considering the fluids from an Eulerian perspective, especially
when considering interface tracking. \Acp{mlm} discretize the domain with
a point cloud that is free to evolve. These points, often referred to as fluid
“particles” carry fluid information with them. This allows the fluid to be
evaluated from a Lagrangian perspective, and thus allows advection to be
handled implicitly by updating fluid particle positions appropriately. In
a multiphase flow context, this provides a natural framework for treating
interface advection in a manner that solely depends on tracking the initial
phase assigned to a particle.

It is common for multiphase \ac{mlm} schemes to adapt multiphase models
originally proposed for mesh-based schemes. One of the most popular multiphase
models for \acp{mlm} remains the color function-based approach first introduced
by Brackbill et al. \cite{brackbill1992}. This work proposed a scalar field
that can be used to distinguish between phases and resolve interfaces, surface
normals, and surface tension forces. The surface normal and surface tension are
resolved using the gradient of the color function. In an Eulerian context, this
field evolves via fluid advection, however, from a Lagrangian perspective, the
color function is a constant particle-level property. This makes introducing
the color function trivial for Lagrangian frameworks and, as a result, has seen
extensive use in various \ac{mlm} schemes including \ac{wc} \ac{sph}
\cite{hu2006, adami2010}, \ac{ic} \ac{sph} \cite{hu2007, hu2009, zainali2013}
and both \ac{wc} \cite{shakibaeinia2012} and \ac{ic} \cite{khayyer2013}
\ac{mps} method.  With \acp{mlm} primarily categorized by their discretization
of differential operators, the choice of scheme influences the resolution of
the color gradient. To improve the accuracy of the interface resolution,
higher-order gradient schemes are often used to resolve the gradients
\cite{adami2010, douillet2019, geara2022, yang2022}, despite these differential
operators being more computationally expensive to evaluate.

When resolving momentum diffusion operators, \ac{sph} schemes typically utilize
operators based on a hybrid \ac{sph}-\ac{fd} Laplacian operators of Morris et
al. \cite{morris1997} or the artificial viscosity-based diffusion operator
originally proposed by Monaghan and Gingold \cite{monaghan1983} as
a stabilizing method. Hu et al. \cite{hu2006} proposed a modified Morris-style
diffusion operator in the context of multiphase flow by basing the hybrid
\ac{sph}-\ac{fd} approximation on the form of the \ac{sph} pressure gradient
form rather than the velocity gradient. Although both Monaghan and Morris-style
diffusion operators have become popular choices for multiphase flow
applications, as highlighted in \cite{ming2017}, Morris-style diffusion
operators are preferred since they can better handle the deformation and
splitting of the multiphase interfaces. 

While both models allow for variations in viscosity, neither resolve viscosity
sensitivities nor account for the associated momentum diffusion \cite{li2020}.
Although the sensitivities are zero for fluids with constant fluid properties,
a discontinuity in viscosity arises across interfaces with different material
properties. As a result, although momentum far from the interface is diffused
appropriately, the surface shear at the interface is only partially recovered
by these momentum diffusion models. Despite this, these momentum diffusion
operators remain the most common choice for multiphase \ac{sph} schemes, even
in modern work \cite{douillet2019, li2020, cheng2022, he2022, yang2022}.

As an alternative to \ac{sph}, the \ac{gfd} method \cite{lanson2008,
lanson2008_2, basic2018} offers an alternative way of constructing an \ac{mlm}
scheme by building the required differential operators from weighted
Taylor-series expansions directly. The differential operator approximations are
of a higher order than classical \ac{sph} and are more  computationally
efficient to compute than the equivalent gradient-correct \ac{sph} operators
\cite{joubert2021}, but, as with gradient-corrected \ac{sph}, this comes at
a  relaxation of the physical symmetries strongly enforced by classical
\ac{sph}. Regardless, \ac{gfd} has been used to simulate both \ac{ic}
\cite{basic2019} and \ac{wc} \cite{joubert2021} fluids and have also shown to
be well-suited for multi-physics applications \cite{joubert2022}.

This work systematically explores and compares different \ac{mlm} diffusion
models to quantify the effect of the viscosity sensitivity on the recovered
solutions. First, the responses of the diffusion operators applied to static
fields are compared. Next, to investigate the effect of sharp viscosity
transitions on the \ac{mlm} operators, the diffusion operators are applied to
a steady-state multiphase flow solution. Finally, a direct comparison between
full 3D simulations of a bubble rising in a fluid is performed. Comparisons
between \ac{gfd} and \ac{sph} momentum diffusion operators are performed on
static fields, but in the interest of limiting the computational requirements
of this study, the full 3D simulation only considers the \ac{gfd} diffusion
operators. As part of this paper's output, a computationally inexpensive
viscosity gradient is proposed. 

\section{Numerical methods}
This work is considered in the context of incompressible multiphase flow and as
such the governing equations of motion are given by the incompressible \ac{nse}:
\begin{align}
    \nabla\cdot\mathbf{u} &= 0, \label{eq:mass_eq} \\
    \rho\frac{d\mathbf{u}}{dt} &= -\nabla p + \nabla\cdot\mathbf{\tau} 
                                + \rho\mathbf{g} + \mathbf{S}, \label{eq:mom_eq}
\end{align}
with $\mathbf{u}$ the velocity, $\rho$ the density, $p$ the pressure,
$\rho\mathbf{g}$ the external body force, $\mathbf{S}$ the interface surface
tension force as discussed in Section \ref{section:multiphase} and
$\mathbf{\tau} = 2\mu\mathbf{T}$ a Newtonian shear stress tensor with $\mu$ the
dynamic viscosity and:
\begin{equation}
    \mathbf{T} = \frac{1}{2}\left[\nabla\otimes\mathbf{u} 
               + (\nabla\otimes\mathbf{u})^T\right], \label{eq:diffusion}
\end{equation}
the symmetric strain rate tensor for incompressible flow. 

It should be noted that \eqref{eq:mom_eq} makes use of the total time
derivative due to the Lagrangian nature of \acp{mlm} automatically treating
advection by updating particle positions. Furthermore, no additional work is
needed to track surface interfaces as the interface update is directly obtained
from the updated fluid particle positions.

Although all material properties are kept constant in this work, both density
and especially dynamic viscosity are allowed to vary between fluids, leading to
discontinuities in the density and viscosity fields at fluid interfaces. As
such, a major focus of this work is set on the diffusion operator at the fluid
interfaces where the viscosity gradient is non-zero.

\subsection{Meshless operators} \label{section:diffops}
Although \ac{mlm} differential operator approximations stem from various
different contexts, the connectivity between points in these meshless schemes
generally follows a kernel weighting approach allowing point-wise information
to locally diffuse in its neighboring area. This work uses a quintic kernel
\cite{morris1997}:
\begin{equation}
    W(\mathbf{r}, h) = w_0
    \begin{cases}
        (3 - q)^5 - 6(2 - q)^5 + 15(1 - q)^5 & \text{for $0 \leq q \leq 1$}\\
        (3 - q)^5 - 6(2 - q)^5               & \text{for $1 <    q \leq 2$}\\
        (3 - q)^5                            & \text{for $2 <    q \leq 3$}\\
        0                                    & \text{otherwise}
    \end{cases}
\qquad\text{with}\qquad q = {\lVert}\mathbf{r}{\rVert} / h,
\end{equation}
with $\mathbf{r}$ a position vector, $3h$ the support radius, $w_0$
a normalization constant and $\lVert\cdot\rVert$ indicating the Euclidean norm.
For \ac{sph} operators in 2D, $w_0 = 7/478\pi$. As a renormalized scheme,
\ac{gfd} operators do not require a normalized kernel and so set $w_0=1$.
A finite support radius is used to limit particle interaction to only its local
neighborhood. The notation $W_{ij} = W(\mathbf{r}_i - \mathbf{r}_j, h)$ is used
to indicate the kernel weighting between particles $i$ and $j$. Furthermore,
this notion is extended to general fields as well using $f_i=f(\mathbf{r}_i)$
to indicate the function value of the field $f$ at the $i^{th}$ particle's
position. Finally, $\mathbf{r}_{ij} = \mathbf{r}_i-\mathbf{r}_j$ is used to
indicates the position of the $i^{th}$ particle relative to the $j^{th}$
particle.

As proposed by Zainali et al. \cite{zainali2013}, jump discontinuities in
material properties are smoothed by applying a Shepard filter
\cite{shepard1968}:
\begin{equation}
    \widetilde{f}_i = \frac{\sum_{j\in I} f_jW_{ij}}{\sum_{j\in I} W_{ij}}, \label{eq:shepard}
\end{equation}
with $f$ either density or viscosity, $I = \left\{ j \in \mathbb{Z}/N\mathbb{Z}
: W_{ij} > 0\right\}$ the indexing set of the $i^{th}$ particle's neighbors and
$\widetilde{\left(\cdot\right)}$ indicating a filtered property.

As required by the multiphase model, the so-called color function $C^n$ is
introduced as:
\begin{equation}
    C^n_i = \frac{\sum_{j\in I^n} W_{ij}}{\sum_{j\in I} W_{ij}},
\end{equation}
with $I^n$ the indexing set of all nodes in the support radius of $i^{th}$
particle and in the $n^{th}$ phase. Since each phase's material properties are
constant and with only two phases  considered in this work, all filtered fields
can be resolved based on one of the color functions:
\begin{equation}
    \widetilde{f}_i = f^0C_i^0 + f^1C_i^1 = f^1 + (f^0 - f^1)C_i^0, \label{eq:color}
\end{equation}
with $f^0$ and $f^1$ the relevant material properties of the first and second
phases, respectively.

The remainder of this section is dedicated to the \ac{gfd} and \ac{sph}
differential operators. It should be mentioned that a full \ac{sph} scheme is
not considered in this work, and so only the momentum diffusion operator is
discussed. Although both \ac{gfd} gradients and Laplacians are discussed in
detail, this section also mainly focuses on the construction of the momentum
diffusion operator.

\subsubsection{GFD differential operators}
At a high level, the \ac{gfd} method constructs meshless differential operators
by kernel weighting finite-difference terms constructed from Taylor-series
approximations. The \ac{gfd} differential operators used in this work are based
on the gradient approximation of Lanson and Vila \cite{lanson2008,lanson2008_2}
and the Laplacian approximation of Basic et al.  \cite{basic2018}. For
a general field $f$ over a $d$-dimensional Euclidean space at the $i^{th}$
particle, these approximations are given by:
\begin{align}
\gfd{\nabla f}_i &= \mathbf{B}_i\cdot\sum_{j\in I} (f_i - f_j) W_{ij} \mathbf{r}_{ij}, \\
\gfd{\nabla^2 f}_i &= 2d\frac{\sum_{j\in I} (f_j - f_i) W_{ij}\left(1 - \mathbf{r}_{ij}\cdot\mathbf{o}_i\right)}
{\sum_{j\in I} \lVert\mathbf{r}_{ij}\rVert^2 W_{ij}\left(1 - \mathbf{r}_{ij}\cdot\mathbf{o}_i\right)},
\end{align}
With:
\begin{align}
\mathbf{B}_i &= \left(\sum_{j\in I} W_{ij} \mathbf{r}_{ij} \otimes \mathbf{r}_{ij}\right)^{-1},\\
    \mathbf{o}_i &= \mathbf{B}_i\cdot\sum_{j\in I} W_{ij} \mathbf{r}_{ij},
\end{align}
the truncation tensor and offset vector, respectively. While this does not
account for all differential operators of the \ac{nse}, the forcing condition
\eqref{eq:mom_eq} can be constructed using these differential operators.
Specifically, the momentum diffusion operator $\nabla\cdot\mathbf{\tau}$ can be
decomposed as $\nabla\mu\cdot(2\mathbf{T}) + \mu\nabla^2\mathbf{u}$. Clearly,
this form only requires gradient and Laplacian approximations.

\Ac{sph} diffusion operators typically build viscosity smoothing into their
differential operators. To allow for a more direct comparison, a \ac{gfd}
diffusion operator following a similar approach is introduced:
\begin{equation}
    \gfd{\mu\nabla^2\mathbf{u}}_i = 2d\frac{\sum_{j\in I} 
    \left(\frac{2\widetilde{\mu}_i\widetilde{\mu}_j}
    {\widetilde{\mu}_i + \widetilde{\mu}_j}\right)
    (\mathbf{u}_i - \mathbf{u}_j)W_{ij}\left(1 - \mathbf{r}_{ij}\cdot\mathbf{o}_i\right)}
    {\sum_{j\in I} \lVert\mathbf{r}_{ij}\rVert^2 W_{ij}
    \left(1 - \mathbf{r}_{ij}\cdot\mathbf{o}_i\right)}.
\end{equation}
There is no appreciable difference in computational cost between the \ac{gfd}
approximation $\gfd{\mu\g^2\mathbf{u}}$ and $\mu\gfd{\g^2\mathbf{u}}$. As such,
the increased computational cost of utilizing the full shear stress divergence
$\gfd{\frac{\g\cdot\mathbf{\tau}}{\rho}}$ is primarily due to its dependency on
$\gfd{\frac{\g\mu}{\rho}}$ and $\gfd{\g\otimes\mathbf{u}}$.

\subsubsection{SPH diffusion operator}
Unlike \ac{gfd} operators, the \ac{sph} differential operators are based on the
analytical derivative of the continuous fields constructed from a 
kernel-weighting process. As such, rather than building differential operators
from weighted finite difference terms, \ac{sph} differential operators are
based on the kernel gradient $\g_i W_{ij} = \frac{\partial W_{ij}}{\partial
\mathbf{r}_i}$.

Although this is typically suitable for gradient approximations, second-order
differential approximations built from analytical second-order kernel
derivatives are generally poor due to a strong sensitivity to particle disorder
with certain particle configurations even leading to non-physical diffusion.
For this reason, \ac{sph} diffusion operators typically make use of a hybrid
\ac{sph}-\ac{fd} approach.

In the context of multiphase flow, the momentum diffusion operator plays an
important role in the evolution of the multiphase interface. As shown in
\cite{ming2017}, the Morris-style diffusion operator of \cite{hu2006} is more
suitable compared to Monaghan-style operators since it allows for more accurate
interface deformation with sharper interfaces. For this reason, the diffusion
operator of \cite{hu2006} as presented in \cite{ming2017} will be used in this
work:
\begin{equation}
    \gfd{\mu\nabla^2\mathbf{u}}_i = \frac{1}{V_i}\sum_{j\in I}
    \left(\frac{2\widetilde{\mu}_i\widetilde{\mu}_j}
    {\widetilde{\mu}_i + \widetilde{\mu}_j}\right)
    \left(V_i^2 + V_j^2\right)
    \left(\md{u}_i - \md{u}_j\right)
    \frac{\md{r}_{ij}\cdot\g_iW_{ij}}{\lVert\mathbf{r}_{ij}\rVert^2},
\end{equation}
with $V_i$ and $V_j$ the volume of the $i^{th}$ and $j^{th}$ fluid particle.
It should be noted that the traditional correction for singular values in the
denominator is omitted since the definition of $I$ ignores cases with particle
distances equal to 0. 

It should be noted that, unlike the \ac{gfd} operators, the \ac{sph} operators
do not apply any kernel corrections. As such, not only is the order of the
approximations lower, but it also significantly suffers from reduced kernel
support at surfaces. For this reason, additional boundary particles are
generated outside the computational domain for all comparisons between \ac{gfd}
and \ac{sph}.

\subsection{Multiphase flow} \label{section:multiphase}
As highlighted in \eqref{eq:mom_eq}, multiphase physics requires the resolution
of a surface tension forcing $\mathbf{S}$ at fluid interfaces. The surface
tension model first proposed by Brackbill et al. \cite{brackbill1992} in
a \ac{fvm} context has been adapted for \acp{mlm}. More specifically, the
gradient of $C^0$ is used to identify surface interfaces and resolve the
surface normal.

The normal direction $\mathbf{N}_i$ is determined as:
\begin{equation}
    \mathbf{N}_i = \gfd{\frac{\g C^0}{\rho}}_i = \mathbf{B}_i\cdot\sum_{j\in I} 
    \frac{C_i^0-C_j^0}{\bar{\rho}_{ij}}W_{ij}\mathbf{r}_{ij}
\end{equation}
with $\bar{\rho}_{ij} = \frac{1}{2}(\widetilde{\rho}_i + \widetilde{\rho}_j)$
the average density.

Following a modified implementation of the scheme proposed by Yang et al.
\cite{yang2020}, interface particles are identified by filtering the color
gradient magnitude. As such, the normal is resolved as:
\begin{equation}
    \hat{\mathbf{n}}_i = 
    \begin{cases}
        \frac{\mathbf{N}_i}{\lVert\mathbf{N}_i\rVert}&
        \text{if}\quad\widetilde{\rho}_i\lVert\mathbf{N}_i\rVert > \epsilon/h\\
        0 & \text{otherwise}
    \end{cases}
\end{equation}
with $\epsilon$ a user-specified parameter to control the aggressiveness of the
filter.  It was found that $\epsilon=0.01$ was suitable for all simulations in
this work.

The surface curvature $\kappa$ is obtained from the divergence of the surface
normal:
\begin{equation}
    \kappa_i = -\gfd{\g\cdot\hat{\mathbf{n}}}_i = \sum_{j\in I\cap I^m} W_{ij}
    \left(\hat{\mathbf{n}}_j-\hat{\mathbf{n}}_i\right)\cdot\mathbf{B}_i\cdot\mathbf{r}_{ij}
\end{equation}
with $I^m = \left\{j \in \mathbb{Z}/N\mathbb{Z} : \lVert\mathbf{n}_j\rVert
> 0\right\}$ the indexing set of all particles bypassing the filter. This
ensures that only particles on the interface contribute to this approximation.

The acceleration due to surface tension force $\mathbf{s} = \mathbf{S}/\rho$ is
then resolved as:
\begin{equation}
    \gfd{\mathbf{s}}_i = -\sigma\gfd{\g\cdot\hat{\mathbf{n}}}_i\gfd{\frac{\g C^0}{\rho}}_i,
\end{equation}
with $\sigma$ the surface tension coefficient. It should be noted that only
particles indexed by $I^m$ have a non-zero $\gfd{\mathbf{s}}_i$. 

Furthermore, the color gradient is reused when determining the viscosity
gradient with effectively no additional computational cost required. By taking
the gradient of \eqref{eq:shepard} applied to the viscosity, then the viscosity
gradient can be determined as:
\begin{equation}
    \gfd{\frac{\g\mu}{\rho}}_i = (\mu^0-\mu^1)\gfd{\frac{\g C^0}{\rho}}_i \label{eq:vis_grad}
\end{equation}
with $\mu^0$ and $\mu^1$ indicating the dynamic viscosity of the first and
second phases, respectively. It should be noted that since velocity is updated
directly, the form of \eqref{eq:vis_grad} is constructed for velocity diffusion
rather than momentum diffusion.  As such, the velocity diffusion operator is
resolved as:
\begin{equation}
    \gfd{\frac{\g\cdot\mathbf{\tau}}{\rho}}_i =
    \gfd{\frac{\g\mu}{\rho}}_i\cdot\gfd{2\mathbf{T}}_i +
    \widetilde{\nu}_i\gfd{\g^2\mathbf{u}}_i,
\end{equation}
with $\widetilde{\nu}_i = \widetilde{\mu}_i/\widetilde{\rho}_i$ the filtered
kinematic viscosity.

\subsection{Integration scheme}
A projection method similar to incompressible \ac{sph} and \ac{mps} is used to
enforce the incompressibility condition \eqref{eq:mass_eq}. An initial update
step is used to introduce the viscous momentum diffusion, surface tension
forces, and other body forces. This results in an unconstrained intermediate
velocity field $\mathbf{u}^*$. A divergence-free velocity field is obtained by
resolving the pressure such that the pressure gradient  suppresses
$\g\cdot\mathbf{u}^*$.

For the $k^{th}$ update step, the intermediate particle velocity is resolved
as:
\begin{equation}
    \mathbf{u}_i^* = \mathbf{u}_i^k + \Delta t\left(\gfd{\mathbf{L}}_i 
                   + \mathbf{g}_i + \gfd{\mathbf{s}}_i\right), \label{eq:vel_update}
\end{equation}
with $\Delta t$ the time-step size and  $\gfd{\mathbf{L}}_i$ the appropriate
velocity diffusion operator. The velocity field is then resolved via the
projection step:
\begin{equation}
    \mathbf{u}_i^{k+1} = \mathbf{u}_i^{*} - \Delta t\gfd{\frac{\g p^{k+1}}{\rho}}_i. \label{eq:vel_proj}
\end{equation}
where $p^{k+1}$ is the pressure field that results in the divergence-free
velocity and:
\begin{equation}
    \gfd{\frac{\g p}{\rho}}_i =  \mathbf{B}_i\cdot\sum_{j\in I}
    \frac{p_i - p_j}{\bar{\rho}_{ij}}W_{ij}\mathbf{r}_{ij}.
\end{equation}

The system is closed by taking the divergence of \eqref{eq:vel_proj}, and
setting $\g\cdot\mathbf{u}_i^{k+1} = 0$ to obtain the \ac{ppe}:
\begin{equation}
    \gfd{\frac{\g^2 p^{k+1}}{\rho}}_i = \frac{1}{\Delta t} \gfd{\g\cdot\mathbf{u}^*}_i.
    \label{eq:ppe} 
\end{equation}

Discretizing the \ac{ppe} results in a sparse large linear system that is
solved iteratively with a \ac{bicg} method with a Jacobi pre-conditioner. The
density field may become noisy when violent mixing takes place resulting in
particles from one phase becoming surrounded by another. To introduce
smoothing, the average density is used to scale each finite-difference term. As
such, the pressure Laplacian of is resolved as:
\begin{equation}
    \gfd{\frac{\g^2 p^{k+1}}{\rho}}_i = 
    2d\frac{\sum_{j \in I}\left(\frac{p_j-p_i}{\bar{\rho}_{ij}}\right)
    W_{ij} \left(1 - \mathbf{r}_{ij}\cdot\mathbf{o}_i\right)}
    {\sum_{j \in I}\lVert\mathbf{r}_{ij}\rVert^2W_{ij}
    \left(1 - \mathbf{r}_{ij}\cdot\mathbf{o}_i\right)}.
\end{equation}
Again, since the material property is constant for individual phases, this
approximation reduces to $\gfd{\nabla^2 p}_i / \rho_i$ for any fluid particle
far from the multiphase interface resulting in the typical single-phase
incompressible \ac{gfd} model.

The particle position is updated according to:
\begin{equation}
    \mathbf{r}_i^{k+1} = \mathbf{r}_i^{k} + 
    \frac{\Delta t}{2}\left(\mathbf{u}_i^{k} + \mathbf{u}_i^{k+1}\right)
\end{equation}

Finally, to address clustering, the anti-clustering algorithm of Xu et al.
\cite{xu2009} is used to shift the particles and create a more uniformly
distributed particle collection.

\section{Results}
\subsection{Verification study}
In this section, the momentum diffusion operators of Section
\ref{section:diffops} are validated. The schemes are applied to Franke's
bivariate function \cite{franke1979}:
\begin{equation}
    \begin{split}
    f(x, y) &= 0.75\exp\left(-\frac{1}{4}(9x-2)^2 - \frac{1}{4}(9y-2)^2\right)\\
            &+ 0.75\exp\left(-\frac{1}{49}(9x+1)^2 - \frac{1}{10}(9y+1)\right)\\
            &+ 0.5\exp\left(-\frac{1}{4}(9x-7)^2 - \frac{1}{4}(9y-3)^2\right)\\
            &- 0.2\exp\left(-(9x-4)^2 - (9y-7)^2\right)
    \end{split}
    \label{eq:franke}
\end{equation}
Since the momentum diffusion operators will be applied to this field, a scalar
field $\eta$ representing the dynamic viscosity is also required.  As such, the
approximations are compared against $\eta\nabla^2f$. First, a constant field
$\eta=1$ is chosen to directly compare the analytical Laplacian and its
meshless approximations. 

\begin{figure}
    \begin{subfigure}{0.49\textwidth}
        \centering
        \includegraphics{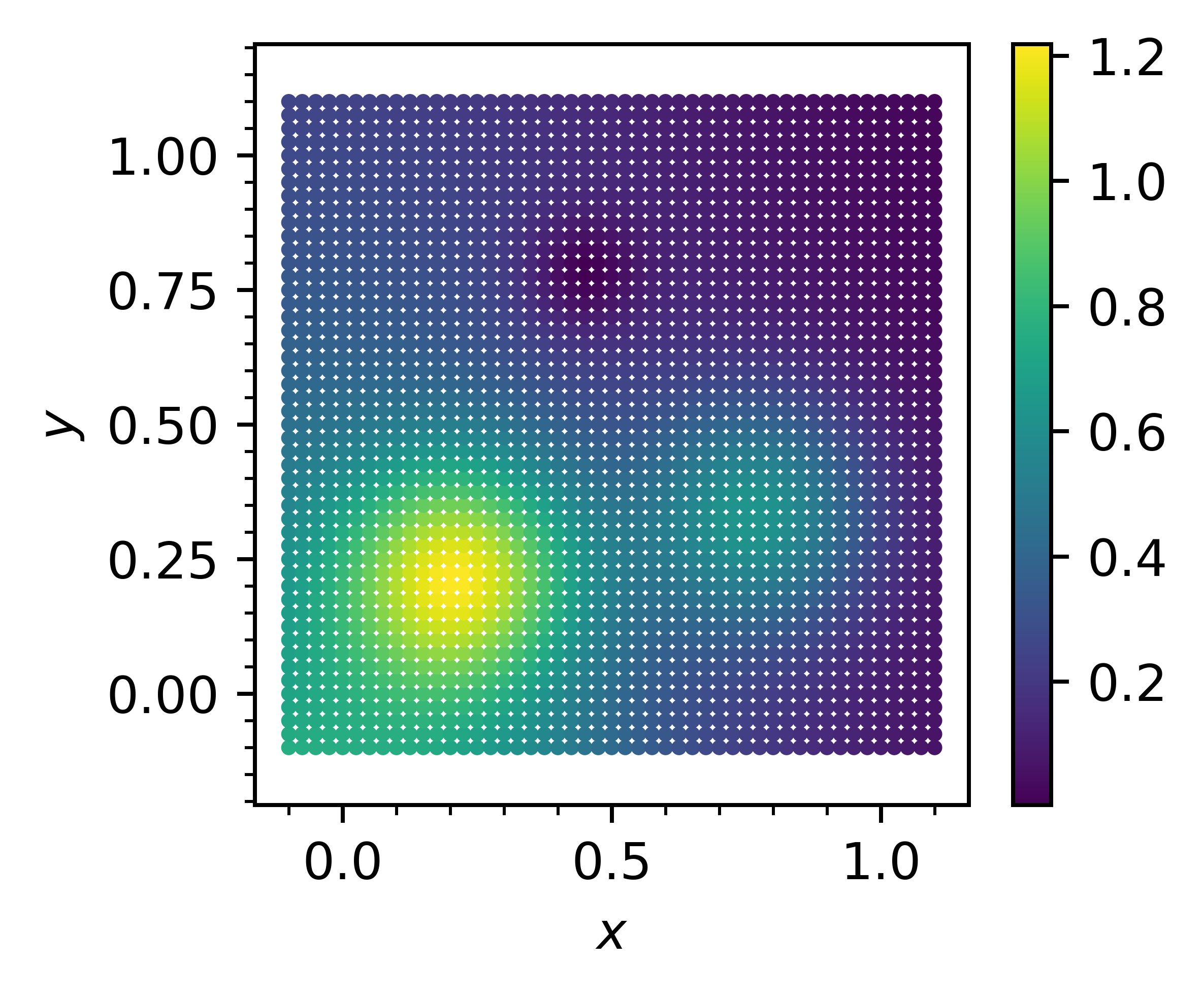}
        \caption{}
        \label{fig:regular}
    \end{subfigure}
    \begin{subfigure}{0.49\textwidth}
        \centering
        \includegraphics{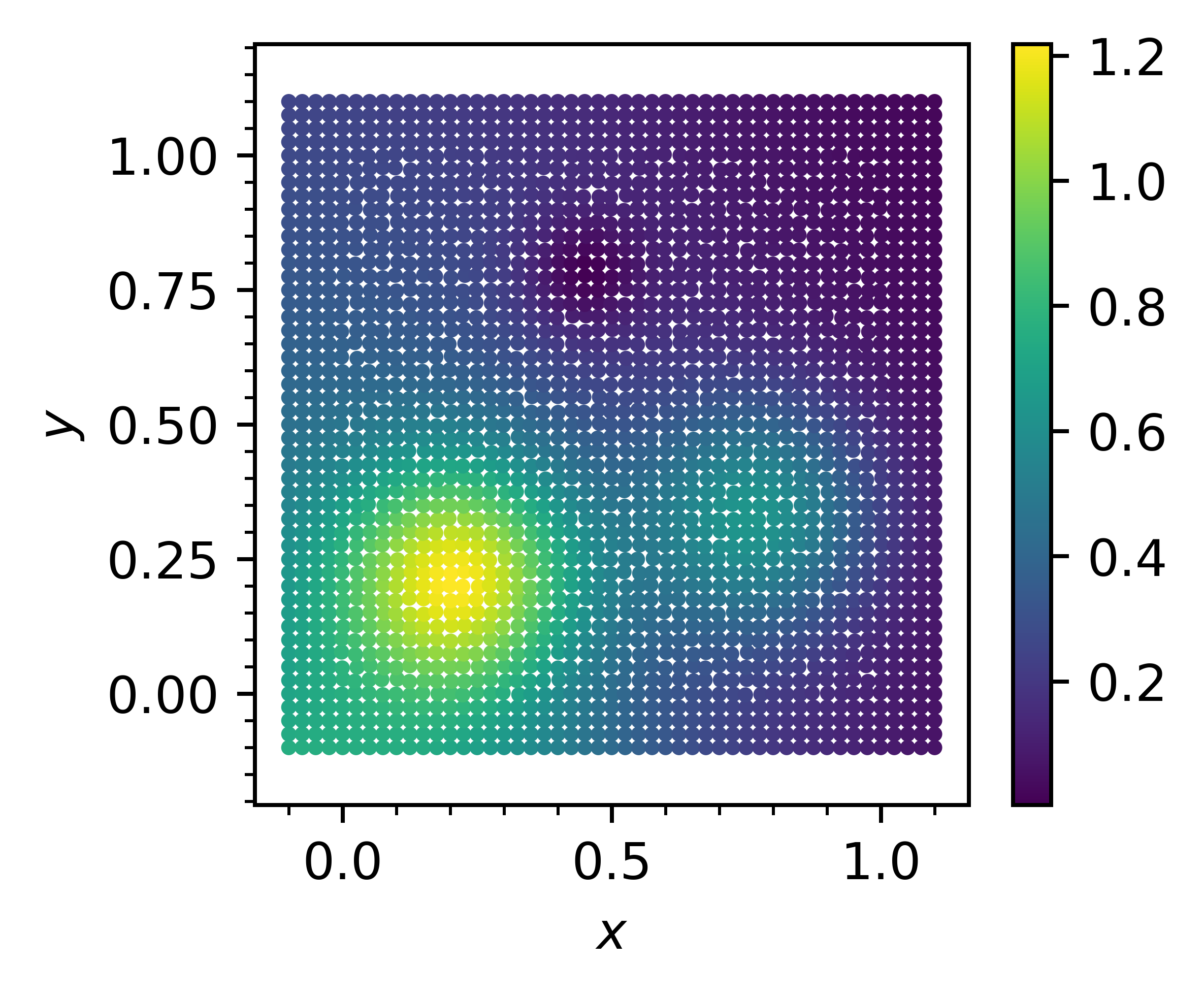}
        \caption{}
        \label{fig:irregular}
    \end{subfigure}
    \caption{Particle configuration for the $41\times41$ (a) regular grid 
    and (b) irregular grid cases.}
\end{figure}

The schemes are tested over both regularly and irregularly spaced points
distributed over $\left\{\left.x,y\in\mathbb{R}^2\right|0\leq x\leq1,0 \leq
y\leq1\right\}$. The irregular configuration is generated by perturbing each
point randomly in both $x$ and $y$ with distances sampled from a uniformly
random distribution with a range of $[-c\delta_0, c\delta_0]$ with $c$ a scale
factor and $\delta_0$ the initial particle spacing. The support radius is given
as $3h = (2.5+2c)\delta_0$. This accounts for the increase in particle
distances due to shifting. All irregular grids in this work use $c=0.1$.
Boundary effects are avoided by generating particles up to $3h$ outside the
computational domain. The internal and boundary nodes for both the regular and
irregular grids with $\delta_0=0.025$ can be seen in Figure \ref{fig:regular}
and Figure \ref{fig:irregular}, respectively. 

\begin{figure}
    \begin{subfigure}{0.49\textwidth}
        \centering
        \includegraphics{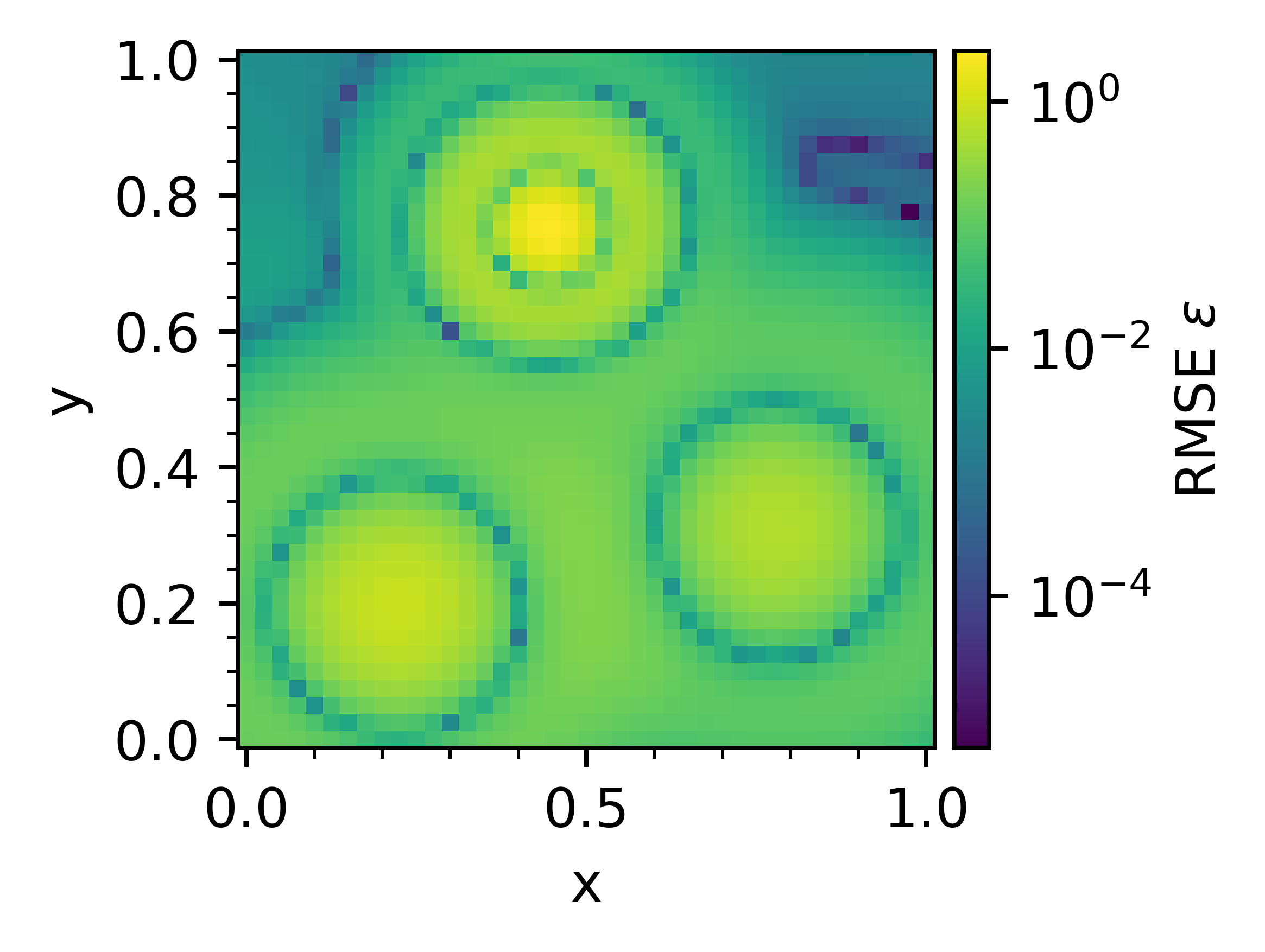}
        \caption{}
    \end{subfigure}
    \hfill
    \begin{subfigure}{0.49\textwidth}
        \centering
        \includegraphics{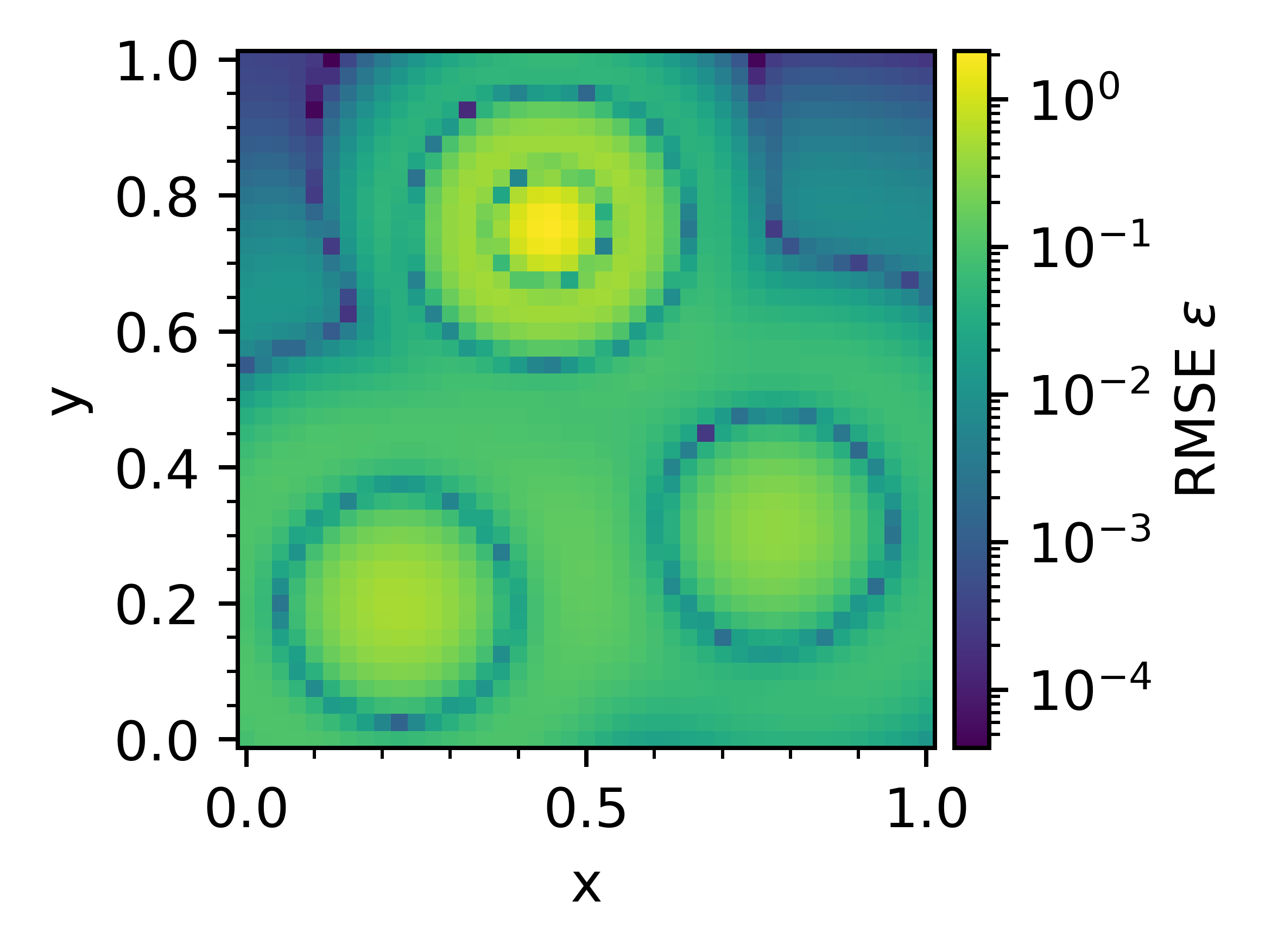}
        \caption{}
    \end{subfigure}
    \hfill
    \caption{Particle level absolute error for the $41\times41$ regular 
    grid case for the (a) \ac{sph} and (b) \ac{gfd} diffusion operators.}
    \label{fig:2d_error_r}
\end{figure}

\begin{figure}
    \begin{subfigure}{0.49\textwidth}
        \centering
        \includegraphics{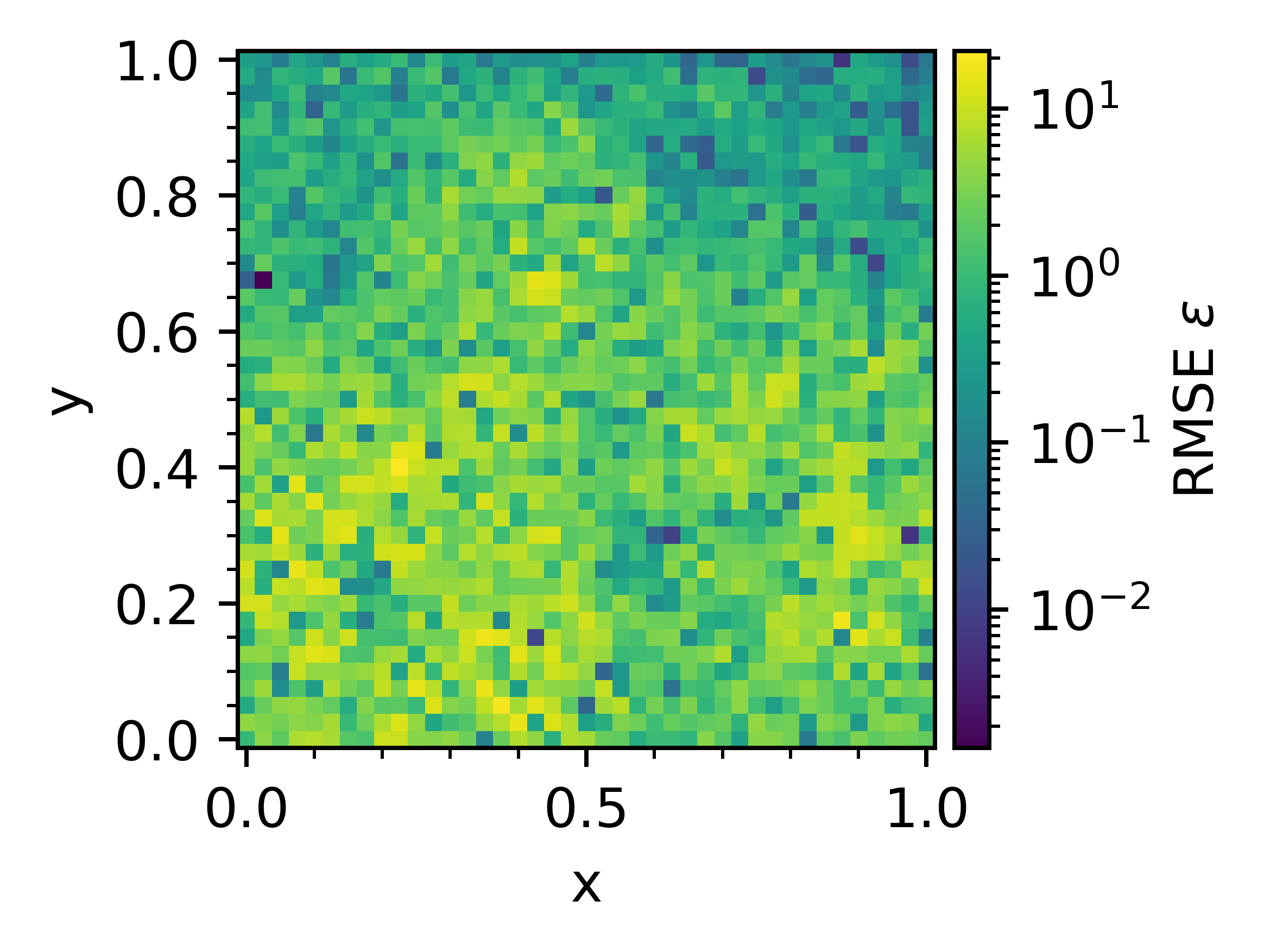}
        \caption{}
    \end{subfigure}
    \hfill
    \begin{subfigure}{0.49\textwidth}
        \centering
        \includegraphics{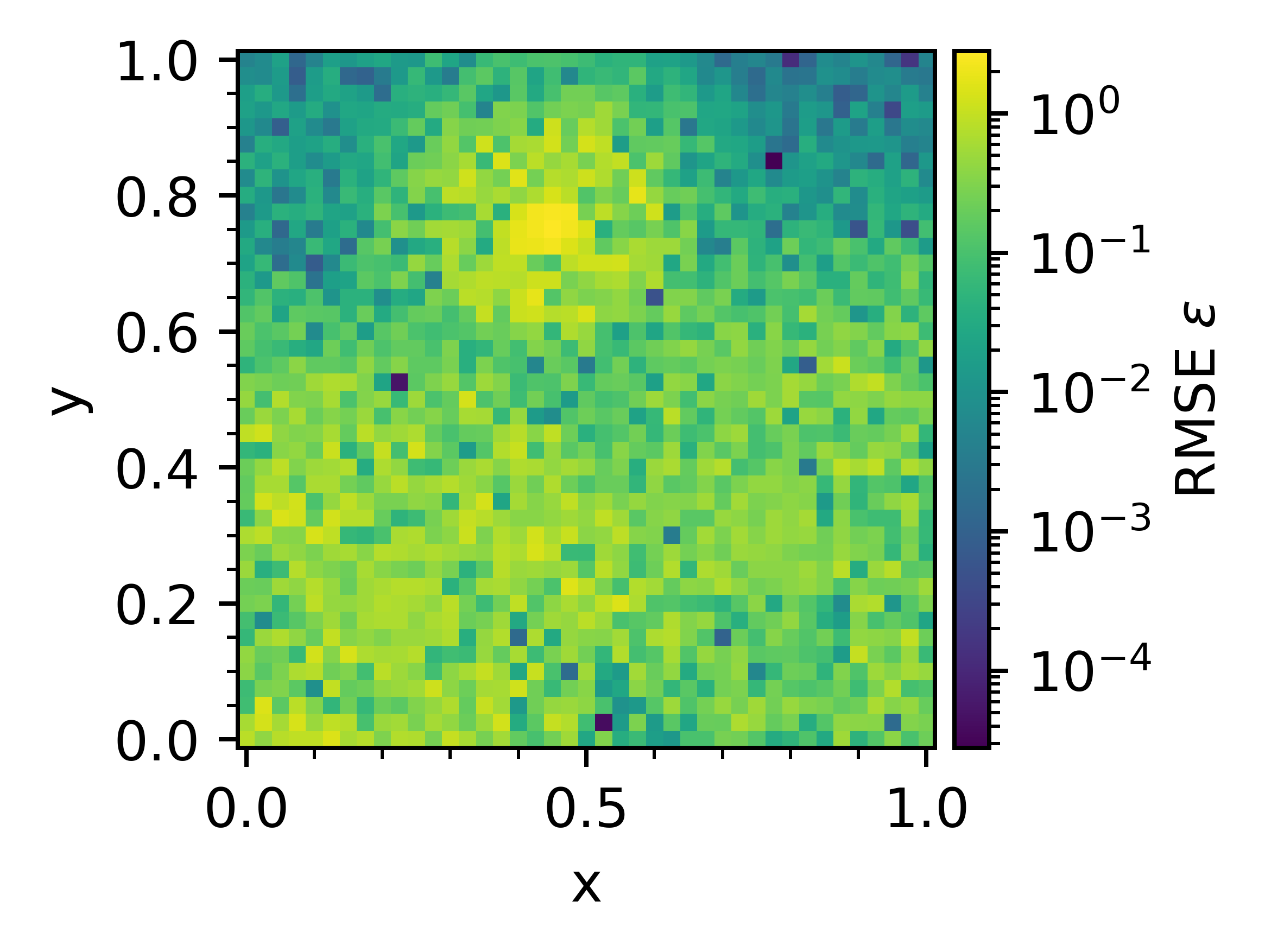}
        \caption{}
    \end{subfigure}
    \caption{Particle level absolute error for the $41\times41$ irregular 
    grid case for the (a) \ac{sph} and (b) \ac{gfd} diffusion operators.}
    \label{fig:2d_error_ir}
\end{figure}

Figure \ref{fig:2d_error_r} shows the absolute error of the \ac{gfd} and
\ac{sph} Laplacian approximations applied to the regular configuration seen in
Figure \ref{fig:regular}. Both the \ac{sph} and \ac{gfd}  operators show
similar qualitative behavior, with the largest errors present around the
function peaks. Although the \ac{gfd} operator has a lower error compared to
the \ac{sph} operator, the errors are in the same order of magnitude.
Similarly, Figure \ref{fig:2d_error_ir} shows the results  for the irregular
configuration. Here it can be seen that the \ac{gfd} errors still align well
with the function peaks. Conversely, the \ac{sph} errors are less dependent on
function shape and primarily driven by the irregularity of the particle
distribution. Compared to the regular grid results, although the \ac{gfd}
results are less accurate on the irregular grid, the operator still produces
errors in the same order of magnitude. The \ac{sph} operator suffers from
a significant error increase when applied to an irregular grid.

\begin{figure}
    \begin{subfigure}{0.49\textwidth}
        \centering
        \includegraphics{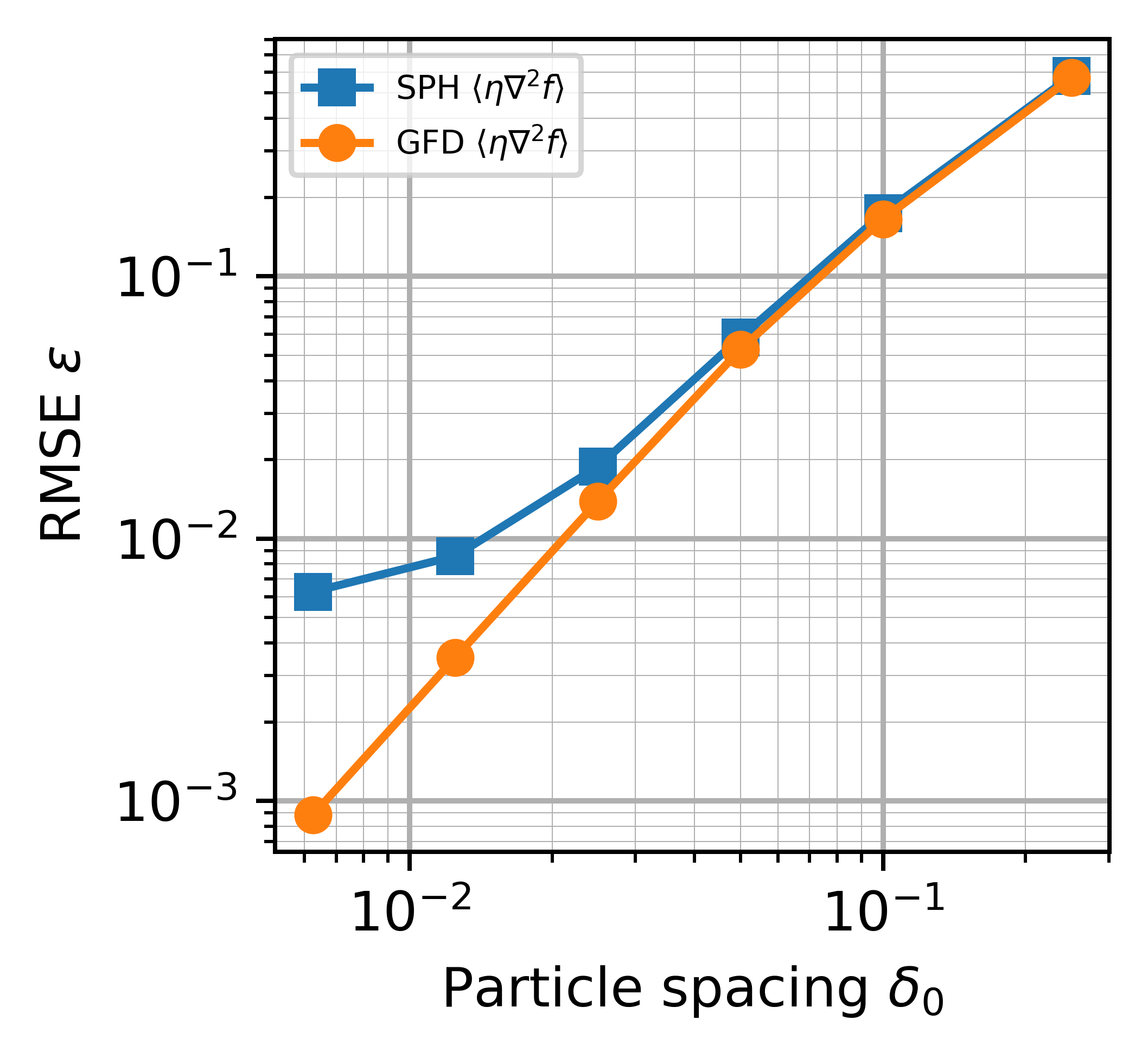}
        \caption{}
    \end{subfigure}
    \hfill
    \begin{subfigure}{0.49\textwidth}
        \centering
        \includegraphics{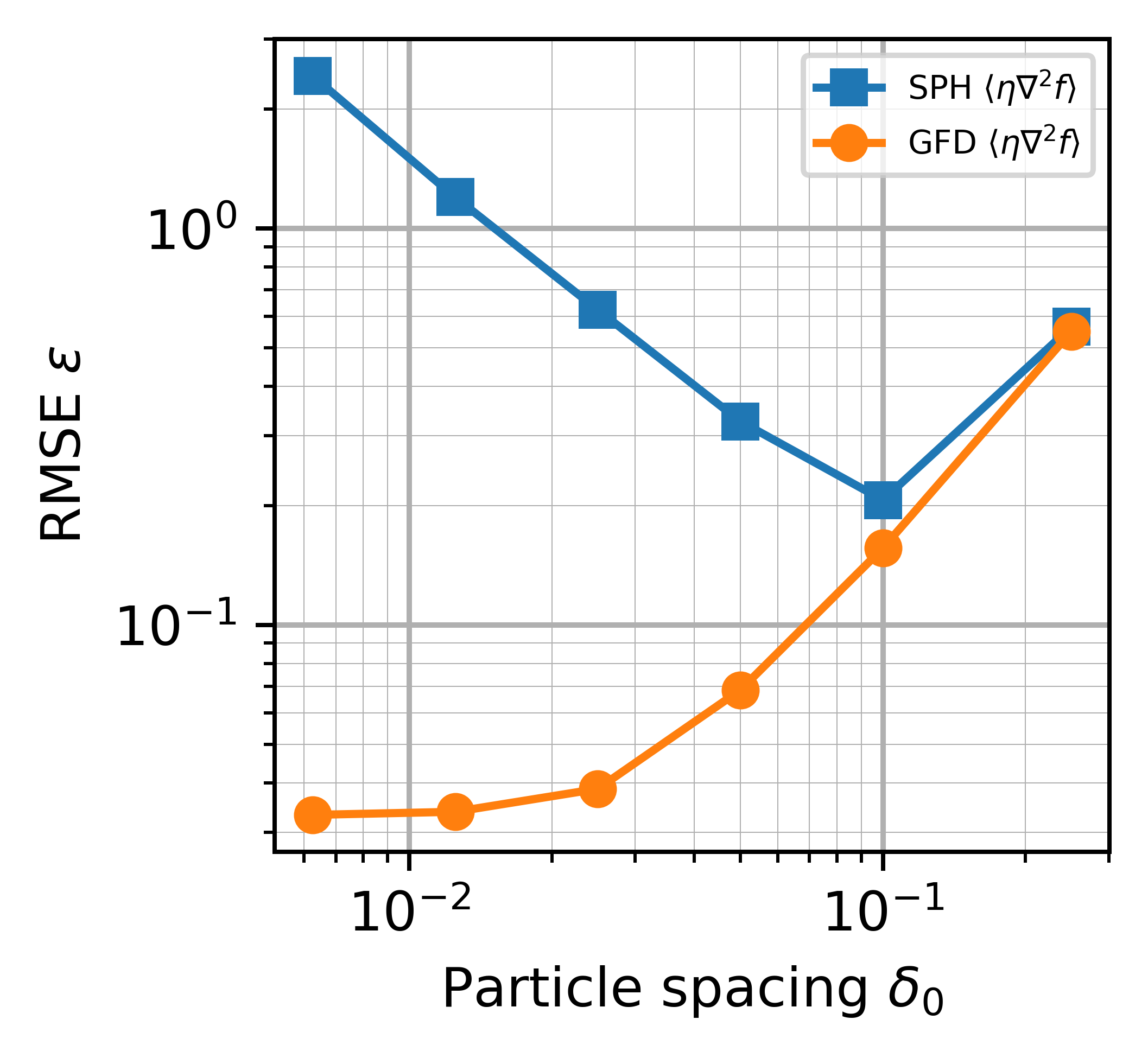}
        \caption{}
    \end{subfigure}
    \caption{\ac{sph} and \ac{gfd} diffusion operators \ac{rmse} on a (a) 
    regular and (b) irregular grid.}
    \label{fig:converge}
\end{figure}

A convergence study for both the \ac{sph} and \ac{gfd} operators on a regular
and irregular grid can be seen in Figure \ref{fig:converge}.  The normalized
\ac{rmse} is defined as: 
\begin{equation}
    \epsilon = \frac{\sqrt{\sum_i\left[\gfd{L(f,\eta)}_i - \eta\g^2f(\mathbf{r}_i)\right]^2}}
    {\sqrt{\sum_i\left[\eta\g^2f(\mathbf{r}_i)\right]^2}}
\end{equation}
with $\gfd{L(f,\eta)}_i$ a momentum diffusion operator. The \ac{rmse} is
plotted against particle size. For the regular spacing, both operators show
similar accuracy initially, but the \ac{sph} operator's rate of convergence
reduces as the number of particles increases while the \ac{gfd} operator shows
a consistent rate of convergence throughout the tested range.  Irregularity can
be seen to have a detrimental effect on both the \ac{sph} and \ac{gfd} error
and rate of convergence. It can be seen that particle irregularity introduces
a limit on the  \ac{rmse} of the \ac{gfd} operator.  Obtaining the same
behavior observed in \cite{basic2018}, the \ac{sph} operator behaves especially
poorly as it sees an increase in the \ac{rmse} with an increase in the number
of particles. This is due to \ac{sph} resolving the Laplacian in an averaged
sense. Regions with high irregularity resolve the Laplacian with large over-
and under-predictions, leading to a large error at the particle level.

\begin{figure}
    \begin{subfigure}{0.49\textwidth}
        \centering
        \includegraphics{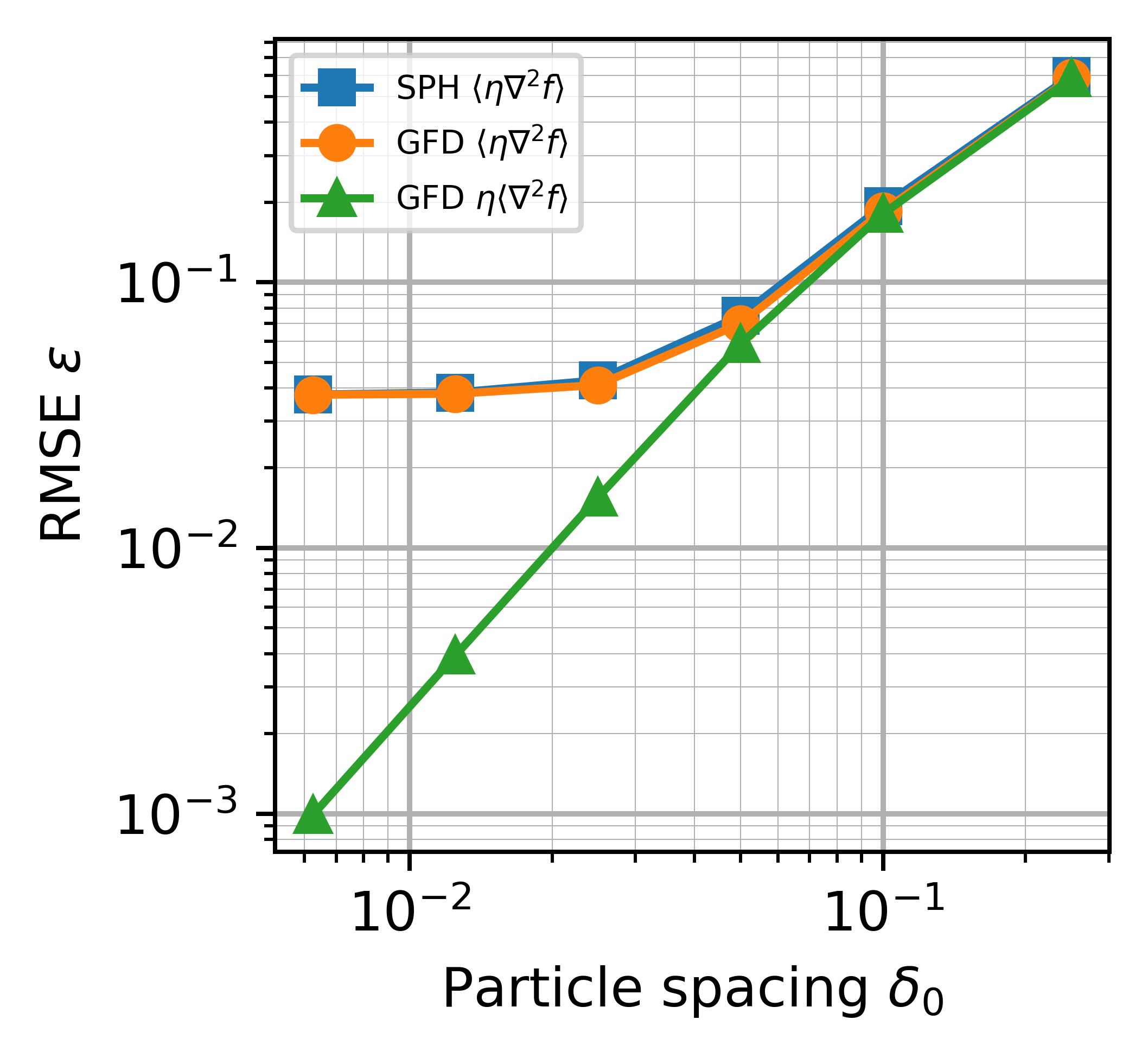}
        \caption{}
    \end{subfigure}
    \hfill
    \begin{subfigure}{0.49\textwidth}
        \centering
        \includegraphics{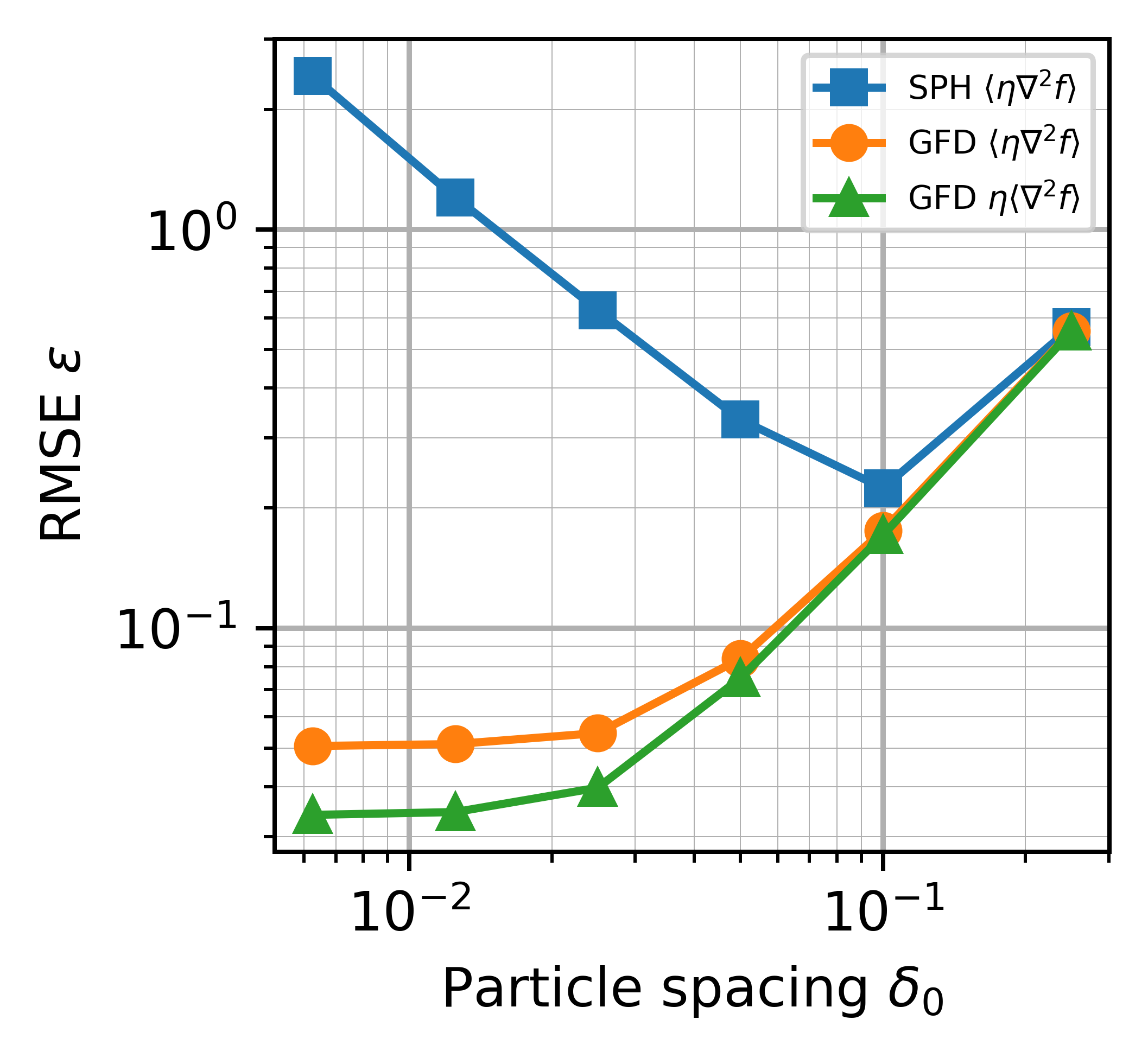}
        \caption{}
    \end{subfigure}
    \caption{\ac{sph} and \ac{gfd} diffusion operators with varying viscosity
    \ac{rmse} on a (a) regular and (b) irregular grid.}
    \label{fig:converge_mu}
\end{figure}

Next, a convergence study is performed with a linearly-varying scale field
given by $\eta(x,y) = 1 + x/2 + y/2$. The results can be seen in Figure
\ref{fig:converge_mu}. Here, along with the \ac{sph} operator, both
$\gfd{\eta\g^2f}$ and $\eta\gfd{\g^2f}$ are  resolved using the \ac{gfd}
operator.

Of course, the \ac{rmse} of $\eta\gfd{\g^2f}$ remains unchanged from the
previous case. Both \ac{gfd} and \ac{sph} $\gfd{\eta\g^2f}$ show similar
behavior on a regularly spaced grid. When comparing results on the irregular
grid, it can be noticed that \ac{sph} performs similarly to the previous case,
showing no significant response to the introduction of a linearly-varying
scalar field. Both \ac{gfd} schemes have similar trends to the previous case,
however, it is found that resolving $\eta\g^2f$ as $\eta\gfd{\g^2f}$ is more
accurate than $\gfd{\eta\g^2f}$. 

\subsection{Multiphase Poiseuille flow}
The next set of results considers the momentum diffusion operators applied to
velocity fields in a multiphase flow environment. Specifically, the
contributions towards momentum diffusion over the interface of the various
schemes are quantified and compared against each other.

\begin{figure}
    \centering
    \includegraphics{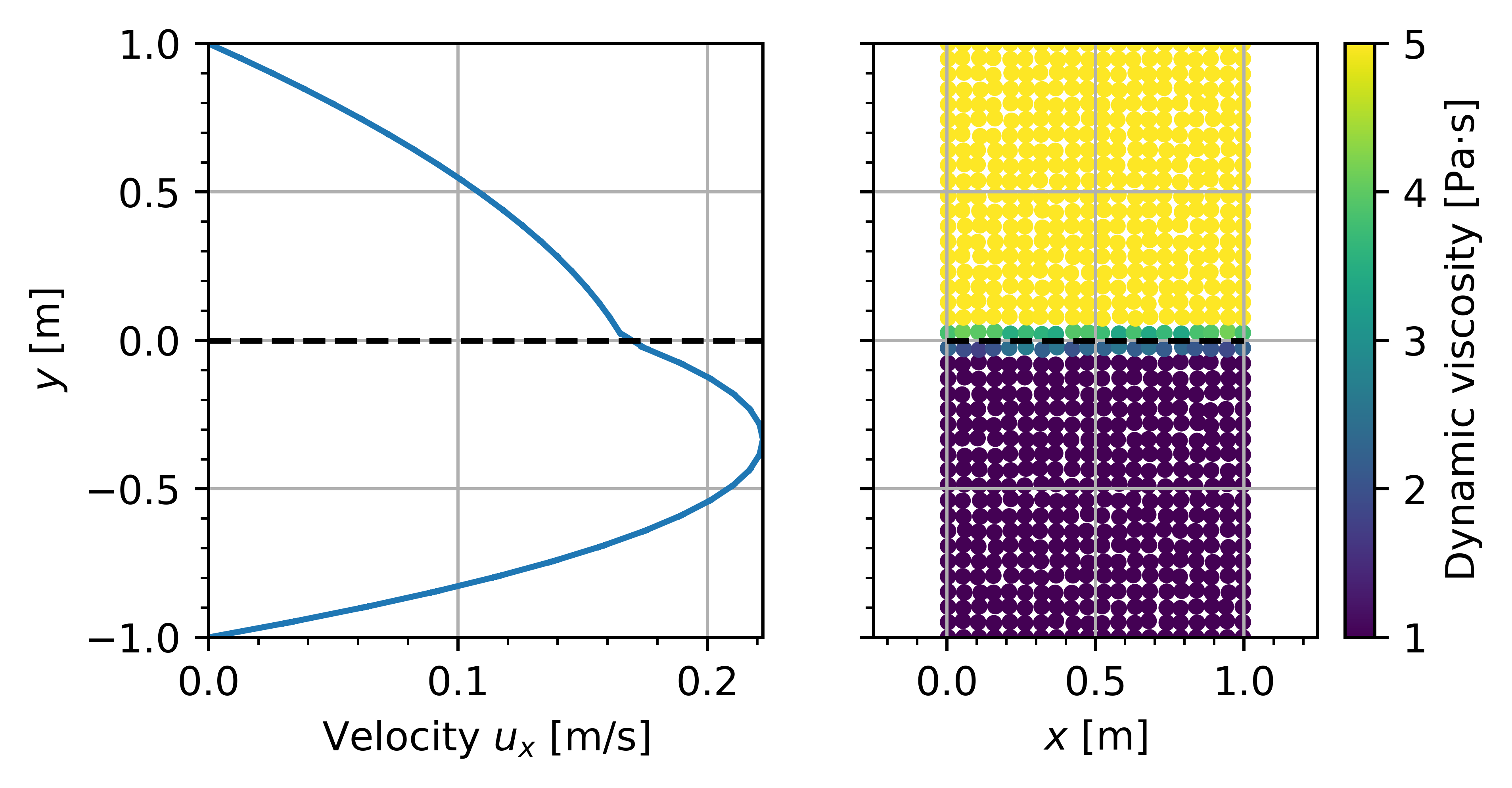}
    \caption{System configuration showing (a) the velocity profile and (b)
    the particle distribution colored by dynamic viscosity for the 
    $\mathcal{X} = 5:1$ case. The dashed line indicates the multiphase 
    interface.}
    \label{fig:poiseuille_2d}
\end{figure}

The numerical experiments are performed on velocity field solutions for
multiphase Poiseuille flow between parallel plates. The velocity field is
assigned according to the analytical results of \cite{bird2002}. Viscosity
ratios $\mathcal{X}$ between 1:1 and 100:1 are considered in this study with
the lower viscosity set to $\mu_L=1\text{Pa}\cdot\text{s}$ for all cases. All
results in this section are considered on an irregular $20\times40$ grid with
$c=0.10$ over the domain $\left\{\left.x,y\in \mathbb{R}^2\right|0\leq
x\leq1,-1\leq y\leq1\right\}$. As with the previous section, boundary
particles are used to extend the computational domain and avoid kernel
truncation. The same grid is used for all cases.  A pressure gradient of
$\frac{dp}{dx}=-1 \text{N}/\text{m}^3$ is used to uniquely determine the
velocity field. Figure \ref{fig:poiseuille_2d} shows the velocity profile and
particle distribution colored by dynamic viscosity for the $\mathcal{X} = 5:1$
case. It should be emphasized that the particle viscosity is obtained from the
color-weighted averaging scheme described in \eqref{eq:color}.

\begin{figure}
    \begin{subfigure}{0.32\textwidth}
        \centering
        \includegraphics{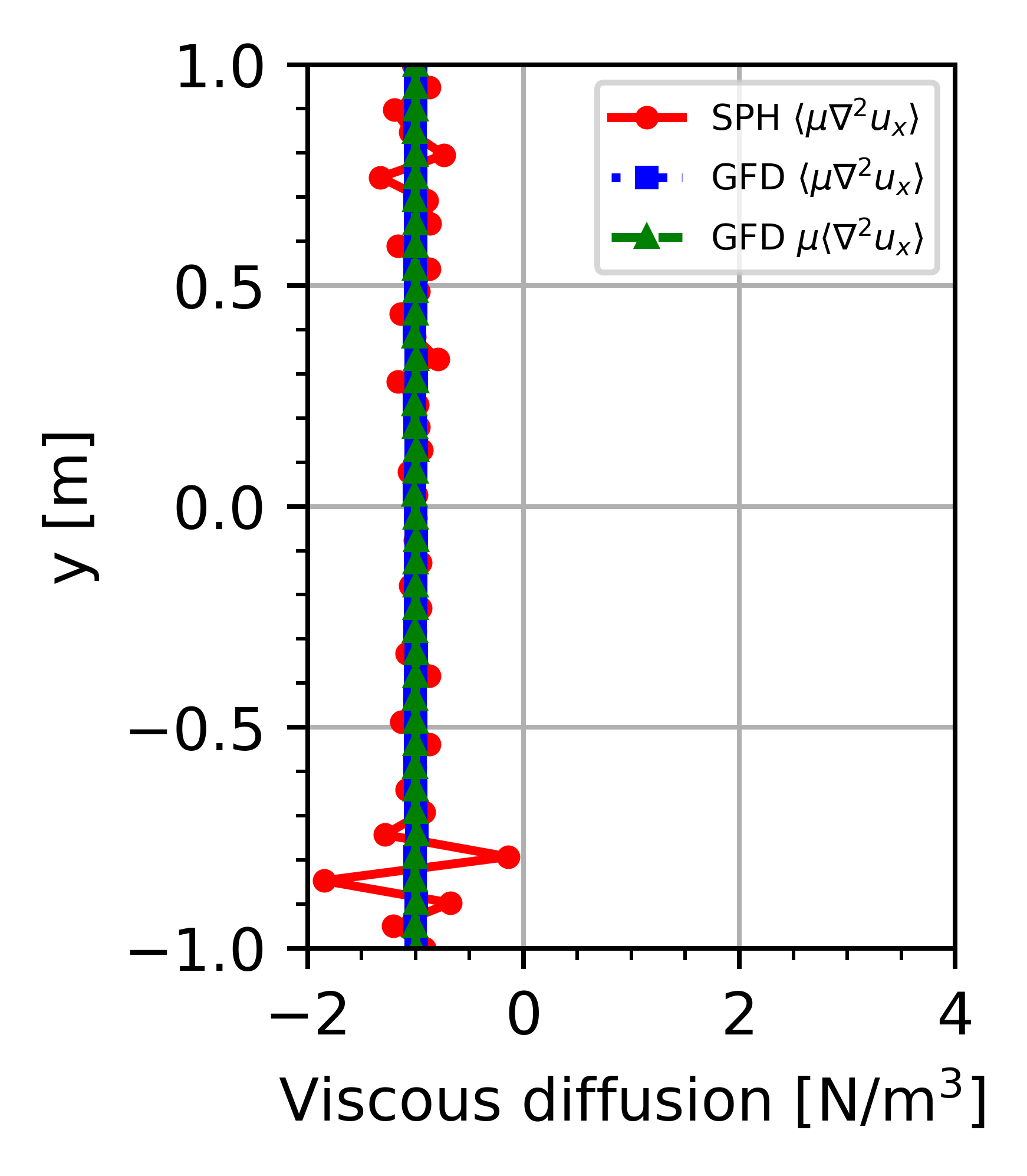}
        \caption{}
        \label{fig:x=1}
    \end{subfigure}
    \hfill
    \begin{subfigure}{0.32\textwidth}
        \centering
        \includegraphics{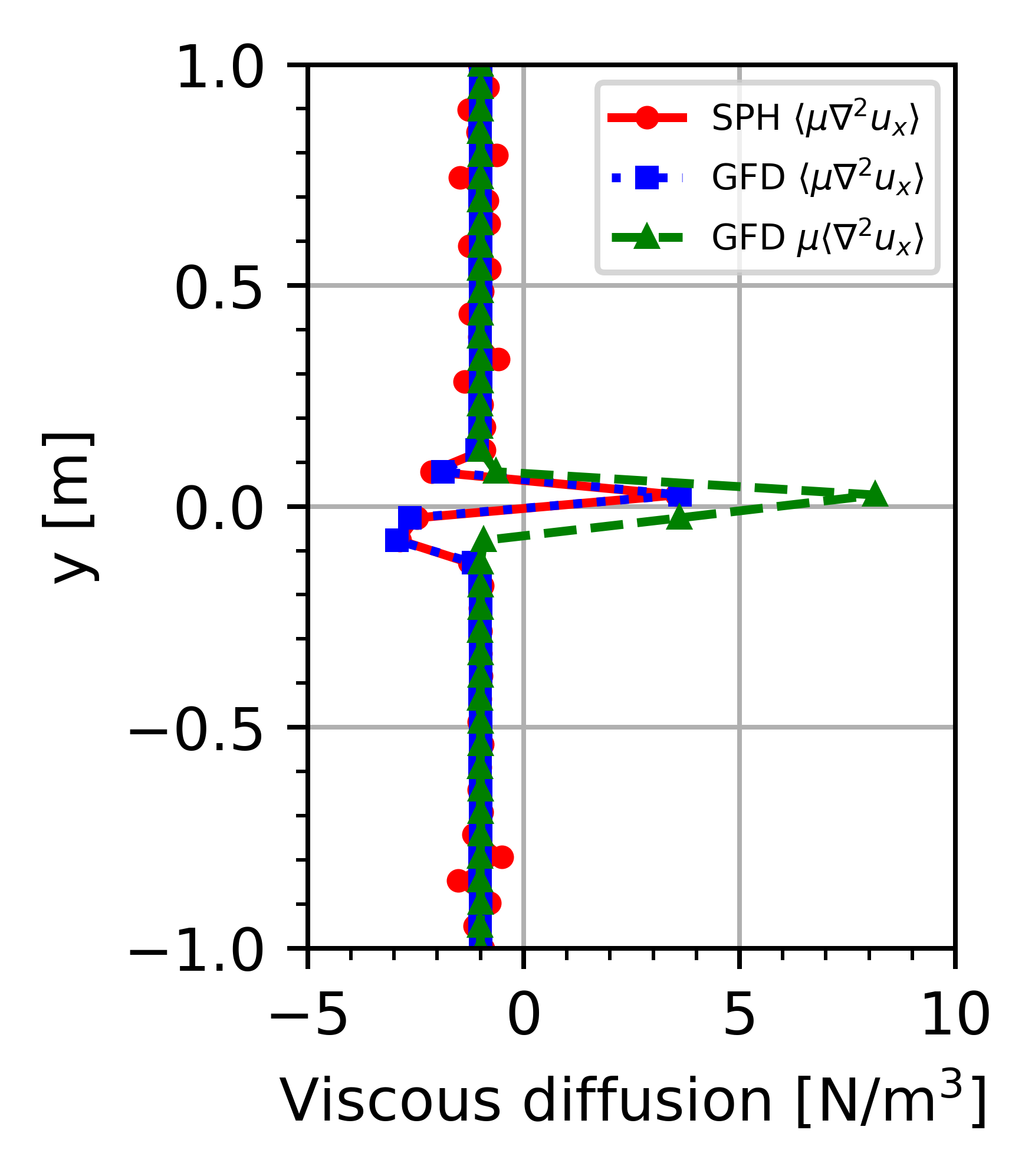}
        \caption{}
        \label{fig:x=5}
    \end{subfigure}
    \hfill 
    \begin{subfigure}{0.32\textwidth}
        \centering
        \includegraphics{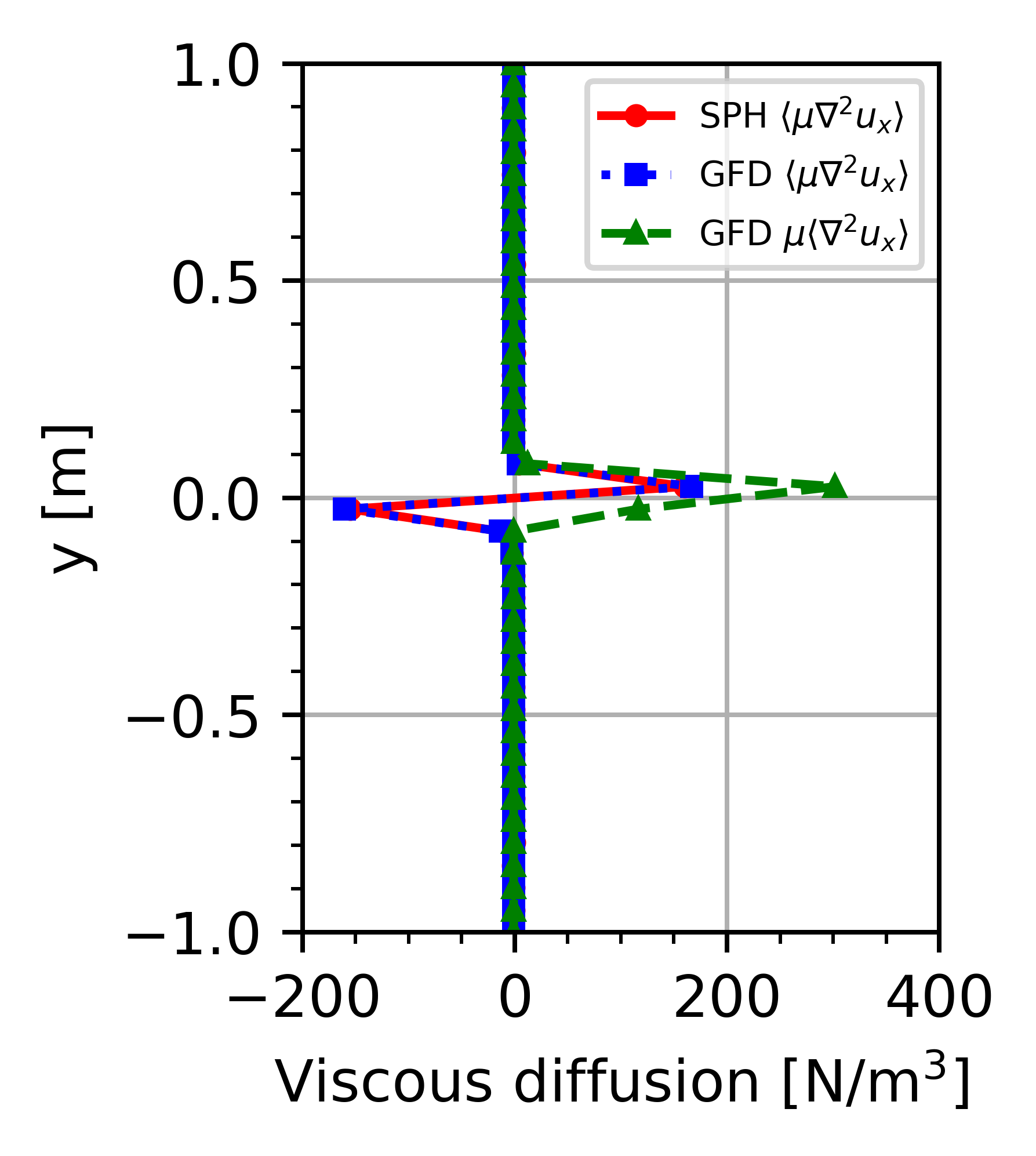}
        \caption{}
        \label{fig:x=100}
    \end{subfigure}
    \caption{Averaged diffusion operator response for (a) $\mathcal{X} = 
    1:1$, (a) $\mathcal{X} = 5:1$ and (c) $\mathcal{X} = 100:1$.}
    \label{fig:lapl}
\end{figure}

Figure \ref{fig:lapl} shows the diffusion operators' responses averaged over
the x-direction for different viscosity ratios. Figure \ref{fig:x=1} serves as
a baseline with the \ac{gfd} and \ac{sph} operators applied to a single-phase
case. Similar to the results of the previous section, it can be seen that the
\ac{sph} results are noisier than the \ac{gfd} results due to grid
irregularity. Figure \ref{fig:x=5} and Figure \ref{fig:x=100} shows results for
$\mathcal{X}=5:1$ and $\mathcal{X}=100:1$, respectively. The behaviour of the
\ac{gfd} approximation $\mu\gfd{\g^2u_x}$ as well as the \ac{gfd} and \ac{sph}
approximations $\gfd{\mu\g^2u_x}$ are provided. These results show that both
the \ac{sph} and \ac{gfd} $\gfd{\mu\g^2u_x}$ diffusion operators have similar
behavior in the multiphase region. It can be seen that $\gfd{\mu\g^2u_x}$
significantly under-predicts the value obtained from the post-multiplied
diffusion operator $\mu\gfd{\g^2u_x}$.

The shear divergence $\gfd{\g\cdot\mathbf{\tau}}$ is resolved with the \ac{gfd}
operators by resolving both $\mu\gfd{\nabla^2\mathbf{u}}$ and
$\gfd{\nabla\mu}\cdot\gfd{2\mathbf{T}}$. To keep the accuracy and order of
operators consistent in this comparison, the operator is only compared against
the \ac{gfd} $\gfd{\mu\g^2\mathbf{u}}$ operator. Furthermore, since the
\ac{sph} operator is shown to behave similarly in the interface region in
a qualitative sense, the findings provide insight for the \ac{sph} operators as
well. It should be noted that $\left(\cdot\right)_x$ is used to indicate the
x-component of a diffusion operator. 

\begin{figure}
    \begin{subfigure}{0.32\textwidth}
        \centering
        \includegraphics{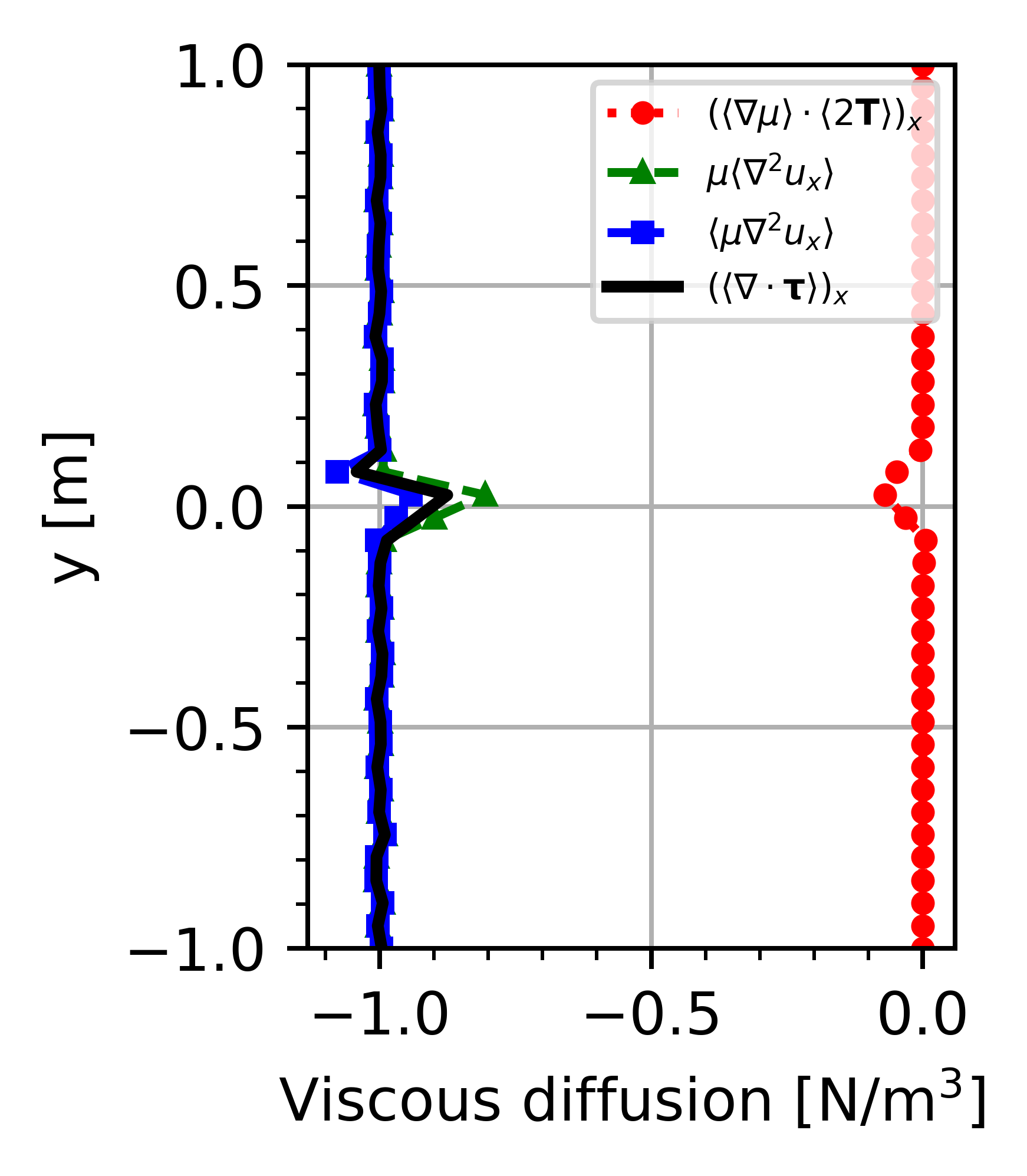}
        \caption{}
        \label{fig:dx=1}
    \end{subfigure}
    \hfill
    \begin{subfigure}{0.32\textwidth}
        \centering
        \includegraphics{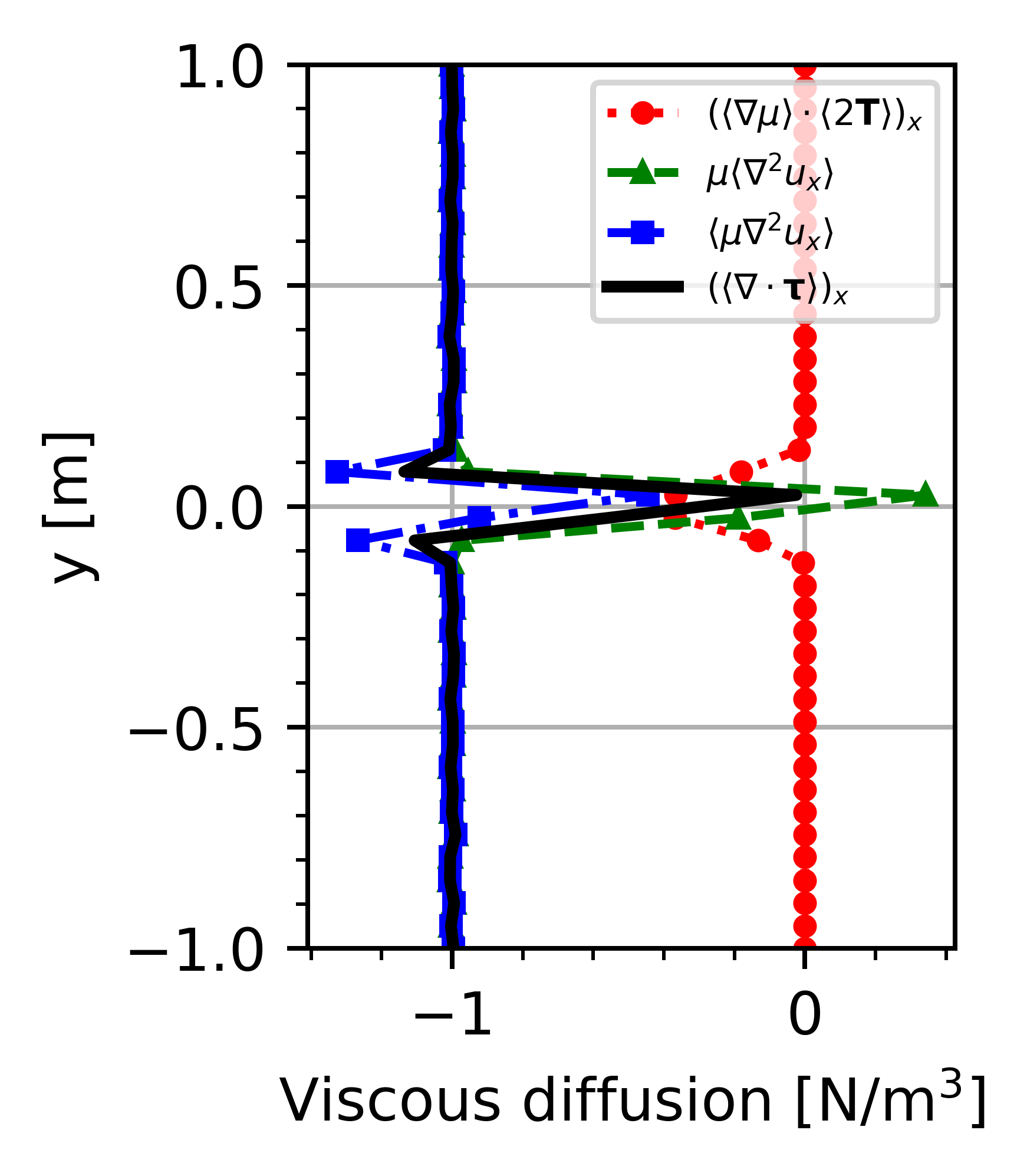}
        \caption{}
        \label{fig:dx=5}
    \end{subfigure}
    \hfill 
    \begin{subfigure}{0.32\textwidth}
        \centering
        \includegraphics{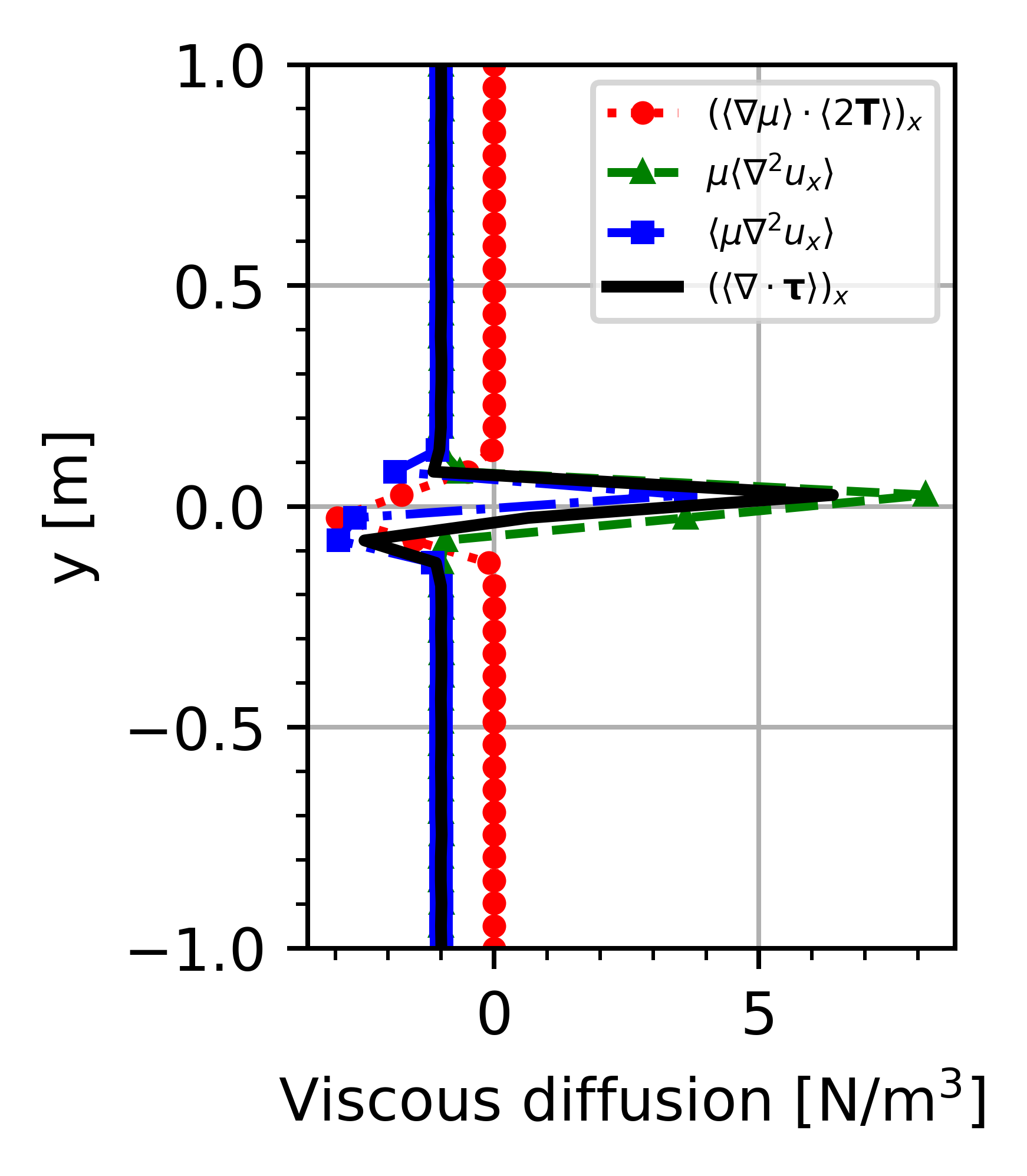}
        \caption{}
        \label{fig:dx=100}
    \end{subfigure}
    \caption{Comparison between \ac{gfd} momentum diffusion operators for 
    (a) $\mathcal{X}=1.3:1$, (a) $\mathcal{X}=2:1$ and (c) $\mathcal{X}=5:1$.}
    \label{fig:tot_mom_comp}
\end{figure}

A comparison for $\mathcal{X}=1.3:1$, $\mathcal{X}=2:1$ and $\mathcal{X}=5:1$
is presented in Figure \ref{fig:tot_mom_comp}. It can be seen that
$\gfd{\nabla\mu} \cdot\gfd{2\mathbf{T}}$ opposes the velocity Laplacian. As
such, at lower viscosity ratios, this balances out the over-prediction of
$\mu\gfd{\nabla^2\mathbf{u}}$ leading to $\gfd{\g\cdot\mathbf{\tau}}$ having
a more similar response to $\gfd{\mu\g^2\mathbf{u}}$. However, as the viscosity
ratio increases, $\gfd{\g\cdot\mathbf{\tau}}$ grows faster than
$\gfd{\mu\g^2\mathbf{u}}$ leading to $\divT$ predicting larger surface shear
forces at the interface.

\begin{figure}
    \centering
    \includegraphics{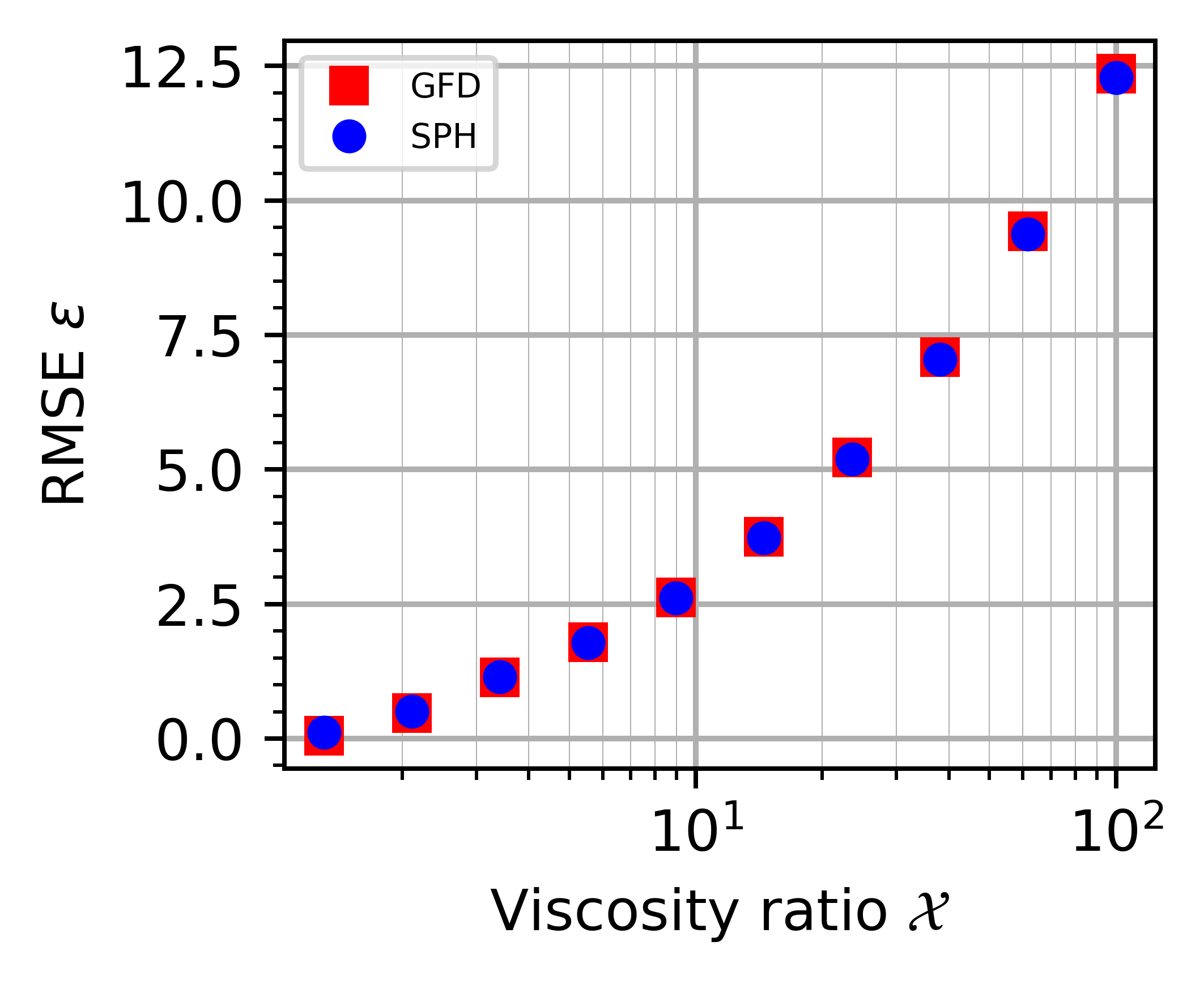}
    \caption{\ac{rmse} of the \ac{gfd} and \ac{sph} $\mixL$ diffusion operators
    over a viscosity ratio range of $\mathcal{X}=1.3:1$ and $\mathcal{X}=100:1$.}
    \label{fig:peak_mu}
\end{figure}

To quantify the difference between the operators, Figure \ref{fig:peak_mu}
shows the \ac{rmse} of the \ac{gfd} and \ac{sph} $\mixL$ operators relative to
the \ac{gfd} $\divT$ operator. Due to the high noise in the \ac{sph} operator,
the \ac{rmse} is only considered over the domain $-3h\leq y\leq3h$. As shown in
the previous results, both \ac{sph} and \ac{gfd} obtain similar results. Here
it is shown that at lower viscosity ratios, all operators predict similar
behavior. However, as $\mathcal{X}$ increases, it can be seen that the
viscosity discontinuity starts to dominate the  surface shear prediction. The
maximum disparity is found at that largest tested viscosity ratio of
$\mathcal{X}=100:1$ where the \ac{rmse} for the \ac{gfd} and \ac{sph} operators
are 12.35 and 12.28, respectively.

\subsection{Rising bubble}
The final section shows the effects that the choice of diffusion model has on
the macroscopic system behavior. In this simulation, an initially spherical
fluid is submerged in a heavier fluid. Buoyancy effects drive the evolution of
the system and are balanced against the drag forces on the bubble surface. This
system is particularly responsive to surface shear since it also partly drives
the bubble's deformation, which further influences the drag characteristics.
Only the $\gfd{\mu\g^2 \mathbf{u}}$ and $\gfd{\g\cdot \mathbf{\tau}}$ diffusion
models resolved with \ac{gfd} operators are compared in this section.

\begin{figure}
    \centering
    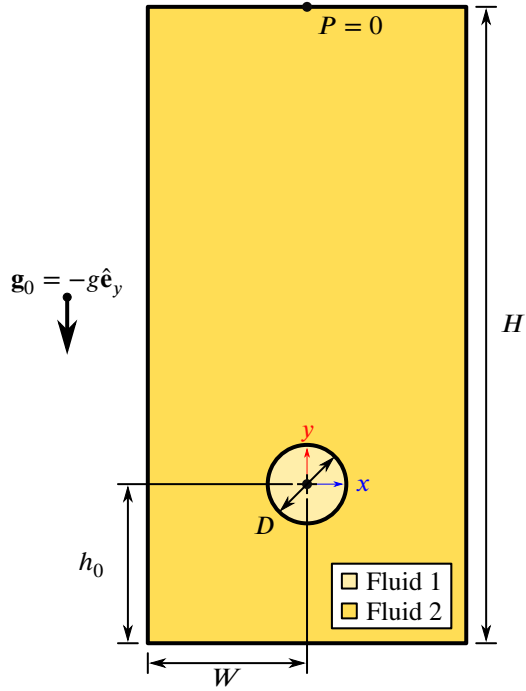
    \caption{Schematic description of the rising bubble case.}
    \label{fig:bubble_scheme}
\end{figure}

Full 3D simulations are performed with a schematic description of the initial
condition shown in Figure \ref{fig:bubble_scheme}. A bubble is initialized
along the centerline of a cylindrical domain. The remaining volume of the
cylinder is filled with a heavier fluid. Full-slip conditions are applied to
the cylinder walls, while no-slip conditions are applied to the cylinder caps.
A point-wise zero-pressure condition is enforced on the top cap at $(0, H-h_0, 0)$
to  allow the pressure field to be uniquely resolved. The density and dynamic
viscosity ratio between the fluids are set to $2:1$ and $100:1$, respectively.
A lower density ratio is chosen to ensure that buoyancy doesn't dominate
viscous effects. The geometry is defined by $H=20D$, $W=2D$, $h_0=2D$ and
$D=1$m. The bubble's density and viscosity are given as $1\text{kg}
/ \text{m}^3$  and $2/875\text{Pa}\cdot s$. The system configuration is
determined by the Reynolds number $Re=\rho_2 D U_g/\mu_2$ and Weber number
$We=\rho_2 D U_g^2 / \sigma$ with $U_g = \sqrt{gD}$ the characteristic
velocity. The gravitational acceleration and surface tension coefficient are
chosen such that $Re=8.75$ and $We=116$. This results in $g=1\text{m}
/ \text{s}^2$ and $\sigma=1/58$N/m. It should be noted that gravity is
incorporated as a buoyancy force and as such, $\mathbf{g}_i = (1-\rho_i/\rho_2)
\mathbf{g}_0$. The initial particle spacing is set to $\delta_0=31.25$mm
leading to a simulation with 4M particles.

Two simulations are performed in this section, with one simulation making use
of the $\mixL$ diffusion model and the other making use of the $\divT$
diffusion model. Besides the diffusion operator, the simulations are identical. 

\begin{figure}
    \begin{subfigure}{0.49\textwidth}
        \centering
        \includegraphics{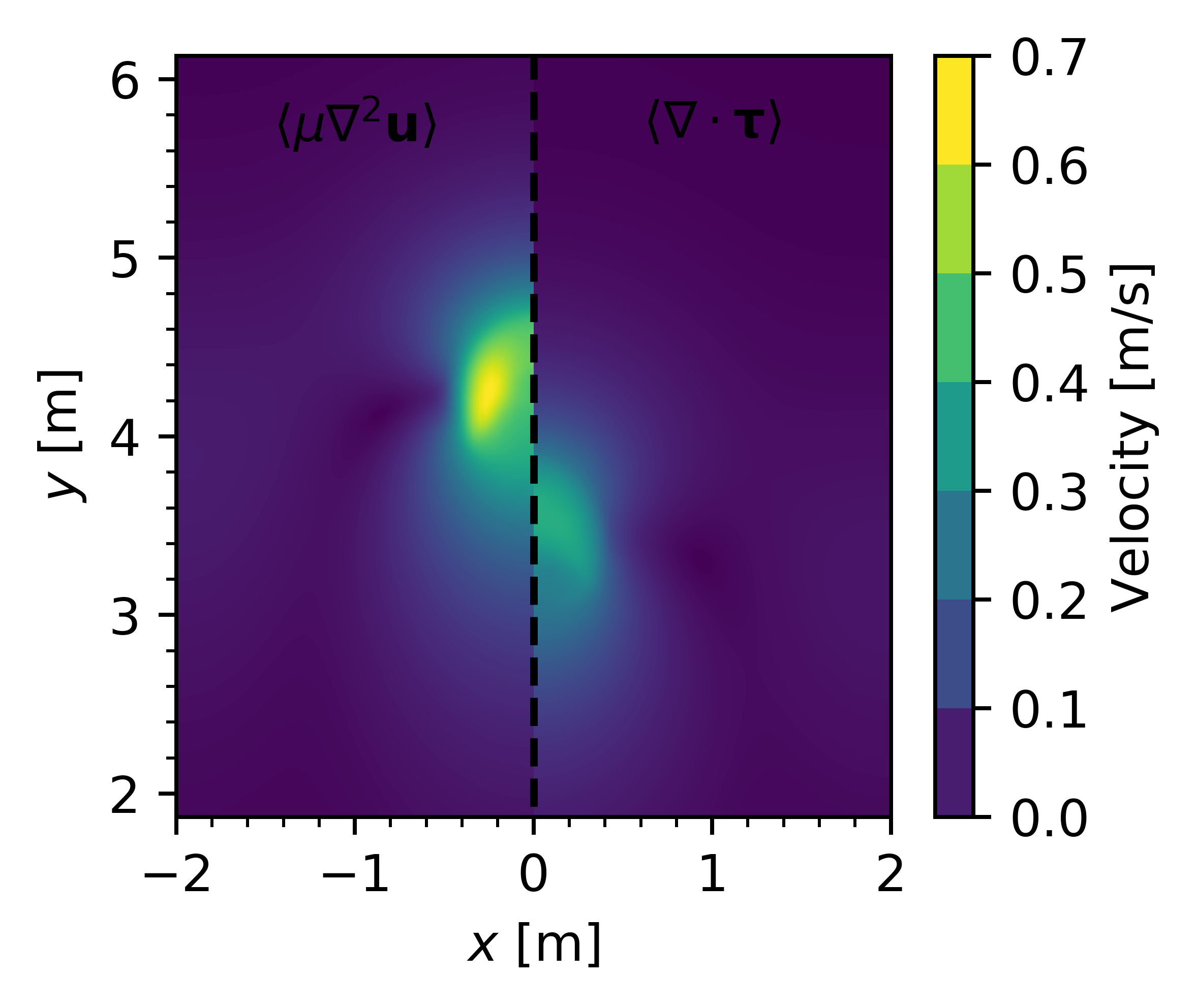}
        \caption{}
    \end{subfigure}
    \hfill 
    \begin{subfigure}{0.49\textwidth}
        \centering
        \includegraphics{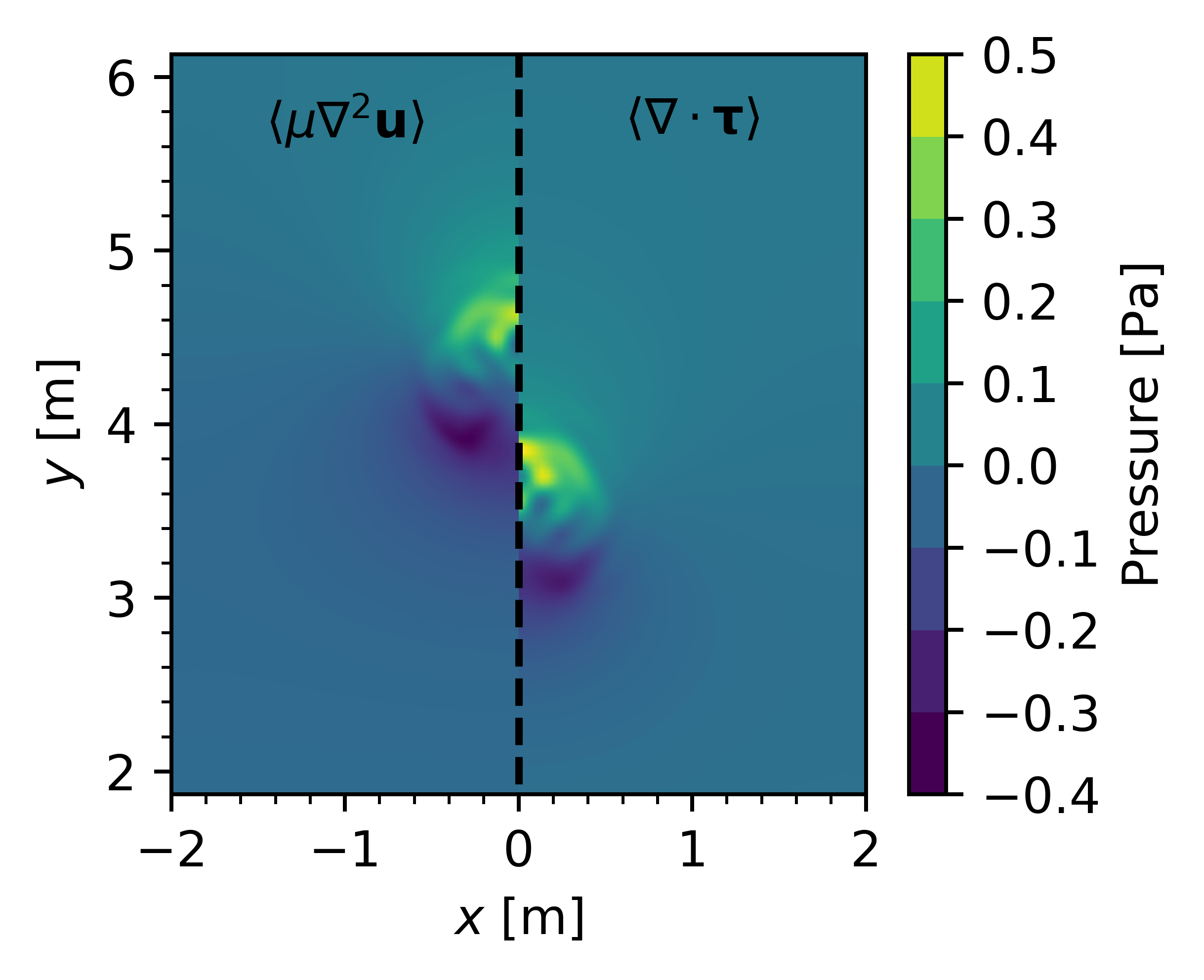}
        \caption{}
    \end{subfigure}
    \caption{Comparison between (a) velocity and (b) pressure fields near the
    bubble at $t=7$s obtained with the $\mixL$ and $\divT$ momentum diffusion
    models.}
    \label{fig:bubble_field}
\end{figure}

Figure \ref{fig:bubble_field} shows the pressure and velocity fields of the
bubbles at $t=7$s. Immediately, it can be seen that the bubble velocity is
significantly lower for the $\divT$ operator. Conversely, the pressure fields
are similar in both magnitude and shape with $\Delta p=0.759$Pa and $\Delta
p=0.782$Pa for the $\mixL$ and $\divT$ cases, respectively. This is a result of
buoyancy being the primary driver behind the pressure response. This indicates
that the difference in bulk bubble kinematics can primarily be attributed to
the viscous diffusion rate.

\begin{figure}
    \centering
    \includegraphics{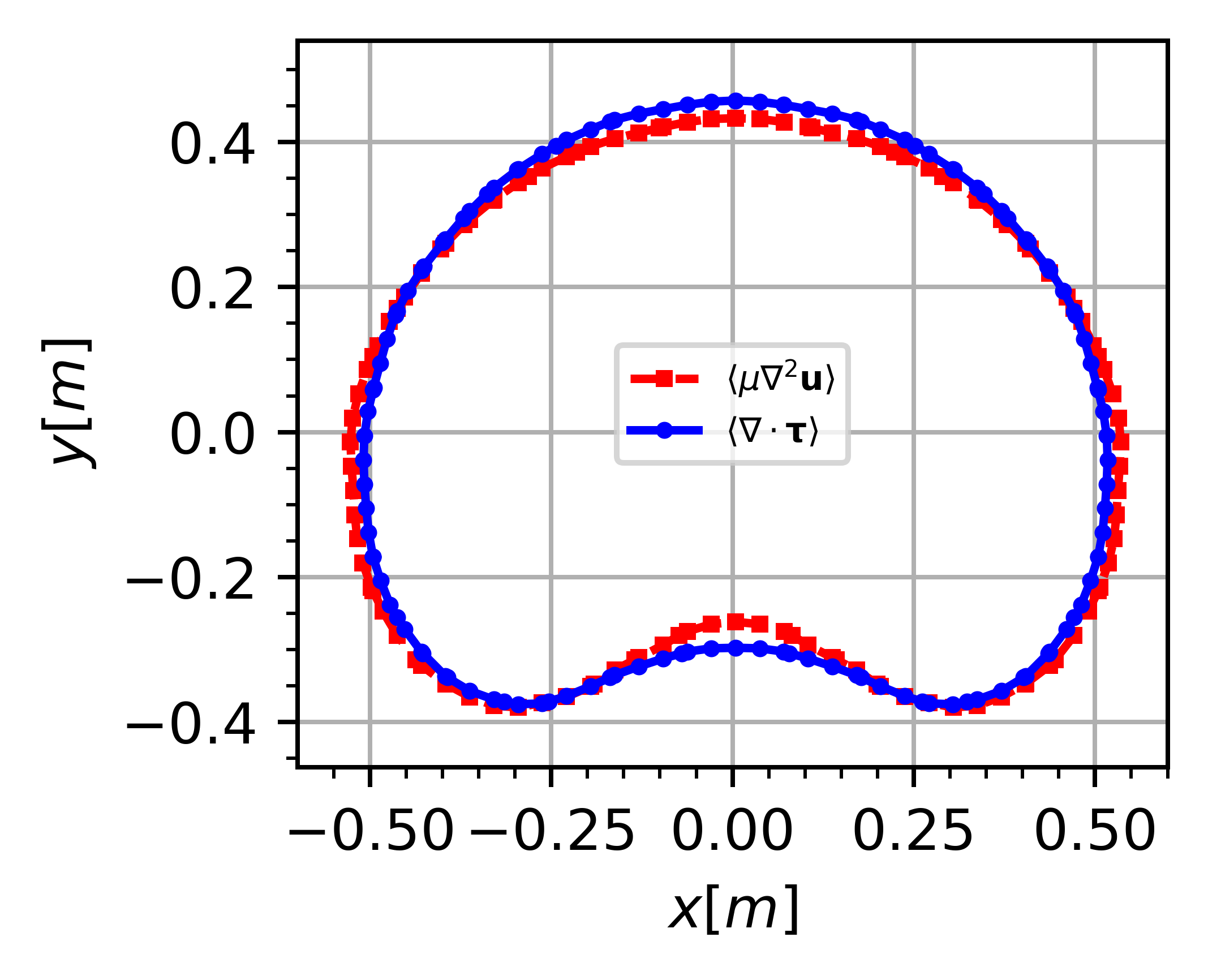}
    \caption{Steady state bubble shapes for the $\mixL$ and $\divT$ diffusion models.}
    \label{fig:bubble_shape}
\end{figure}

The next set of results compares the steady-state bubble shape as seen in
Figure \ref{fig:bubble_shape}. The shape is obtained by applying a Shepard
filter to the color field $C^0$ and resolving the contour at $C^0=0.5$. It can
be noted that the bubble shape is qualitatively similar, although subtle
differences can be seen. Specifically, the $\mixL$ case predicts a slightly
wider and flatter bubble. Furthermore, the bottom surface has a more
exaggerated profile. These differences in bubble shape follow a similar trend
observed in \cite{hua2007} when increasing the density ratio and thus
increasing the ascent rate while fixing $Re=8.75$, $We=116$ and
$\mathcal{X}=100:1$. The effects are not as pronounced as \cite{hua2007},
likely due to the similar densities between phases.

\begin{figure}
    \centering
    \includegraphics{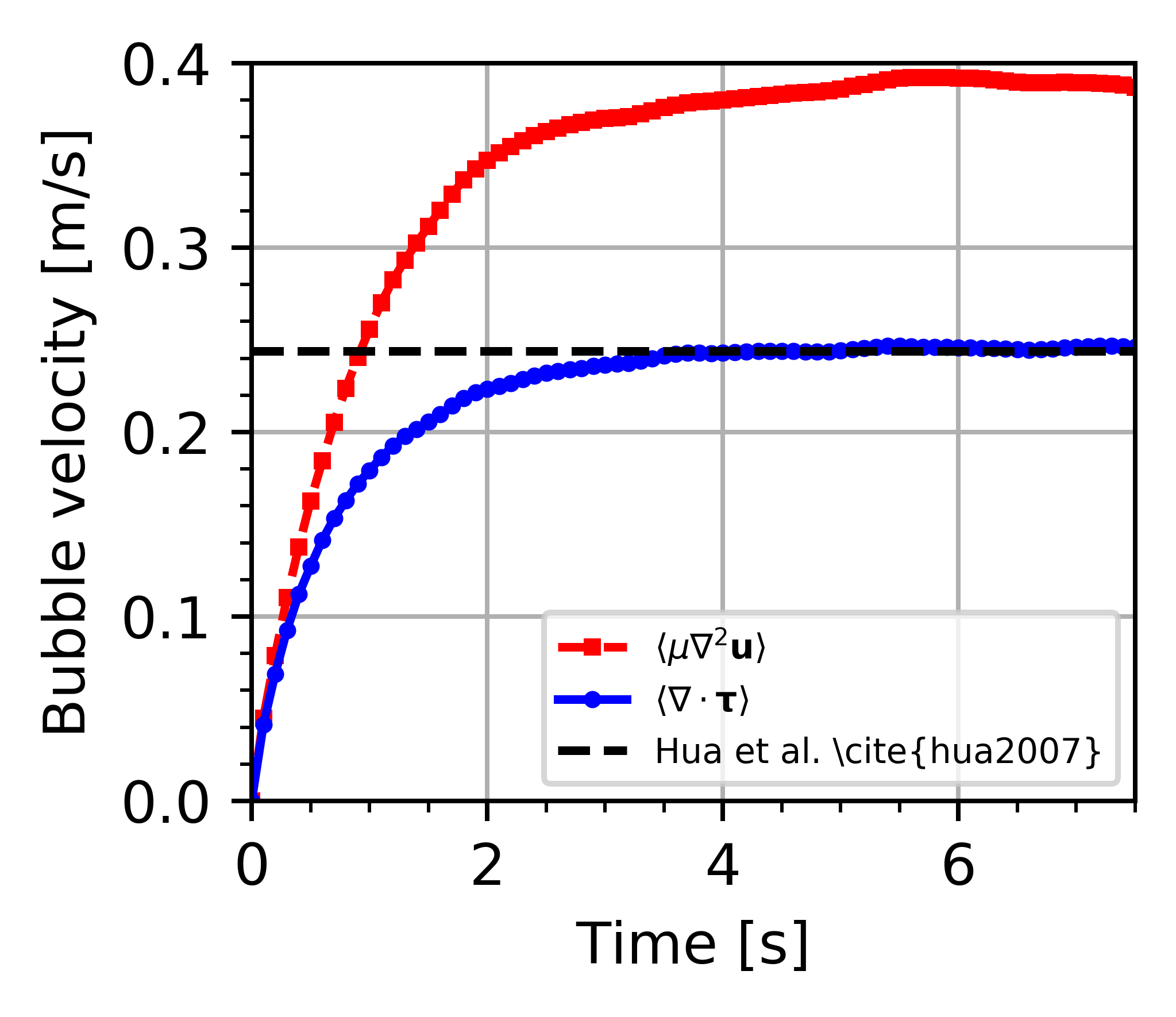}
    \caption{Time evolution of the bubble's bulk vertical velocity for both the
    $\mixL$ and $\divT$ momentum diffusion models.}
    \label{fig:bubble_vel}
\end{figure}

Figure \ref{fig:bubble_vel} shows the average vertical velocity of the bubble.
As shown in the previous results, the $\mixL$ diffusion operator under-predicts
the surface shear. This leads to the $\mixL$ case to predict a higher terminal
velocity with a difference of 57.2\% relative to the $\divT$ case. When
compared to the terminal velocity obtained in \cite{hua2007} using a \ac{fvm},
the $\divT$ diffusion operator predicts similar bubble kinematics.  This is
expected as \cite{hua2007} appropriately resolved the fluid momentum diffusion
as $\nabla\cdot\mathbf{\tau}$.

\section{Conclusion}
This work presents a comparison between \ac{mlm} momentum diffusion models for
multiphase flow. The common momentum diffusion operators for \ac{sph} and
\ac{gfd} are compared to a \ac{gfd} operator that includes the effects of
viscosity sensitivity at multiphase interfaces. A computationally efficient
model for the viscosity gradient based on the color gradient is proposed as
part of this work.

It was shown that although the \ac{gfd} operator was less susceptible to
irregular spacing due to it being first-order accurate, both the \ac{sph} and
\ac{gfd} $\mixL$  momentum diffusion operators behave similarly at multiphase
interfaces. Specifically, both under-estimate surface shear with the error
growing as the viscosity ratio increases. 

The significance of this effect is demonstrated by simulating the evolution of
a bubble submerged in a heavy fluid. The reduced surface shear resulted in
a significant over-prediction of the bubble's terminal velocity despite
resolving similar pressure conditions. Specifically, the ascent velocity is
over-predicted by 57.2\%when modelling momentum diffusion as $\gfd{\mu\g^2
\mathbf{u}}$ compared to $\gfd{\g\cdot\mathbf{\tau}}$ at a density ratio of
$2:1$ and a viscosity ratio of $100:1$.

This indicates that the viscosity gradient cannot be ignored when constructing
the momentum diffusion operator for multiphase flow, especially when viscous
effects are relatively strong. As such, this work recommends that multiphase
\ac{gfd} solvers make use of the proposed $\gfd{\g\cdot\mathbf{\tau}}$
diffusion model. Furthermore, due to the similarity between the \ac{sph} and
\ac{gfd} momentum diffusion operators highlighted in this work, the proposed
diffusion operator should be explored in a \ac{sph} context as well.

\section{References}
\bibliographystyle{elsarticle-num}
\bibliography{bib/bibio}

\end{document}

%% file: bubble_scheme.pdf_tex
\begingroup%
  \makeatletter%
  \providecommand\color[2][]{%
    \errmessage{(Inkscape) Color is used for the text in Inkscape, but the package 'color.sty' is not loaded}%
    \renewcommand\color[2][]{}%
  }%
  \providecommand\transparent[1]{%
    \errmessage{(Inkscape) Transparency is used (non-zero) for the text in Inkscape, but the package 'transparent.sty' is not loaded}%
    \renewcommand\transparent[1]{}%
  }%
  \providecommand\rotatebox[2]{#2}%
  \newcommand*\fsize{\dimexpr\f@size pt\relax}%
  \newcommand*\lineheight[1]{\fontsize{\fsize}{#1\fsize}\selectfont}%
  \ifx\svgwidth\undefined%
    \setlength{\unitlength}{255.18056574bp}%
    \ifx\svgscale\undefined%
      \relax%
    \else%
      \setlength{\unitlength}{\unitlength * \real{\svgscale}}%
    \fi%
  \else%
    \setlength{\unitlength}{\svgwidth}%
  \fi%
  \global\let\svgwidth\undefined%
  \global\let\svgscale\undefined%
  \makeatother%
  \begin{picture}(1,1.01670343)%
    \lineheight{1}%
    \setlength\tabcolsep{0pt}%
    \put(0,0){\includegraphics[width=\unitlength,page=1]{bubble_scheme.pdf}}%
    \put(0.66246602,0.37298686){\color[rgb]{1,0.06666667,0.04705882}\makebox(0,0)[t]{\lineheight{1.25}\smash{\begin{tabular}[t]{c}$y$\end{tabular}}}}%
    \put(0.74528317,0.29316631){\color[rgb]{0,0,1}\makebox(0,0)[t]{\lineheight{1.25}\smash{\begin{tabular}[t]{c}$x$\end{tabular}}}}%
    \put(0,0){\includegraphics[width=\unitlength,page=2]{bubble_scheme.pdf}}%
    \put(0.72598112,0.96915846){\color[rgb]{0,0,0}\makebox(0,0)[t]{\lineheight{1.25}\smash{\begin{tabular}[t]{c}$P=0$\end{tabular}}}}%
    \put(0,0){\includegraphics[width=\unitlength,page=3]{bubble_scheme.pdf}}%
    \put(0.54490218,0.0051836){\color[rgb]{0,0,0}\makebox(0,0)[t]{\lineheight{1.25}\smash{\begin{tabular}[t]{c}$W$\end{tabular}}}}%
    \put(0.96722794,0.52829399){\color[rgb]{0,0,0}\makebox(0,0)[t]{\lineheight{1.25}\smash{\begin{tabular}[t]{c}$H$\end{tabular}}}}%
    \put(0,0){\includegraphics[width=\unitlength,page=4]{bubble_scheme.pdf}}%
    \put(0.59927552,0.22997581){\color[rgb]{0,0,0}\makebox(0,0)[t]{\lineheight{1.25}\smash{\begin{tabular}[t]{c}$D$\end{tabular}}}}%
    \put(0,0){\includegraphics[width=\unitlength,page=5]{bubble_scheme.pdf}}%
    \put(0.3441338,0.17716976){\color[rgb]{0,0,0}\makebox(0,0)[t]{\lineheight{1.25}\smash{\begin{tabular}[t]{c}$h_0$\end{tabular}}}}%
    \put(0,0){\includegraphics[width=\unitlength,page=6]{bubble_scheme.pdf}}%
    \put(0.806341,0.14754349){\color[rgb]{0,0,0}\makebox(0,0)[t]{\lineheight{1.25}\smash{\begin{tabular}[t]{c}Fluid 1\end{tabular}}}}%
    \put(0,0){\includegraphics[width=\unitlength,page=7]{bubble_scheme.pdf}}%
    \put(0.806341,0.10345706){\color[rgb]{0,0,0}\makebox(0,0)[t]{\lineheight{1.25}\smash{\begin{tabular}[t]{c}Fluid 2\end{tabular}}}}%
    \put(0,0){\includegraphics[width=\unitlength,page=8]{bubble_scheme.pdf}}%
    \put(0.30829502,0.59523605){\color[rgb]{0,0,0}\makebox(0,0)[t]{\lineheight{1.25}\smash{\begin{tabular}[t]{c}$\mathbf{g}_0 = -g\hat{\mathbf{e}}_y$\end{tabular}}}}%
  \end{picture}%
\endgroup%